\documentclass[longauth]{aa9}  

\usepackage{ltxtable}
\usepackage{supertabular}
\usepackage{graphicx}
\usepackage{rotating}
\usepackage{pdflscape}
\usepackage{color}
\usepackage{natbib}
\usepackage{amssymb}
\usepackage{lscape}
\usepackage{txfonts}
\bibpunct{(}{)}{;}{a}{}{,} 

\def\logg{$\log$~g}
\def\teff{T$_{\mathrm{eff}}$}

\def\kms{km~s$^{-1}$}

\def\A{$\rm\AA$}

\begin{document} 

\title{The Gaia-ESO Survey: the inner disk, intermediate-age open cluster Trumpler 23}
   
\titlerunning{The Gaia-ESO Survey: open cluster Trumpler 23}

\author{J.~C. Overbeek\inst{1}, E.~D. Friel\inst{1}, P. Donati\inst{2,3}, R. Smiljanic\inst{4}, H. R. Jacobson\inst{5}, D. Hatzidimitriou\inst{6,7}, E.~V. Held\inst{8}, L. Magrini\inst{9}, A. Bragaglia\inst{2}, S. Randich\inst{9}, A. Vallenari\inst{8}, T. Cantat-Gaudin\inst{8,10}, G. Tautvai\u{s}ien\.{e}\inst{11}, F. Jim\'{e}nez-Esteban\inst{12,13}, A. Frasca\inst{14}, D. Geisler\inst{15}, S. Villanova\inst{15}, B. Tang\inst{15}, C. Mu\~{n}oz\inst{15}, G. Marconi\inst{16}, G. Carraro\inst{16}, I. San Roman\inst{17}, A. Drazdauskas\inst{11}, R. \u{Z}enovien\.{e}\inst{11}, G. Gilmore\inst{18}, R.~D. Jeffries\inst{19}, E. Flaccomio\inst{20}, E. Pancino\inst{9,21}, A. Bayo\inst{22}, M.~T. Costado\inst{23}, F. Damiani\inst{20}, P. Jofr\'e\inst{18,24}, L. Monaco\inst{25}, L. Prisinzano\inst{20}, S.~G. Sousa\inst{26}, S. Zaggia\inst{8}}

\institute{Department of Astronomy, Indiana University, Bloomington, IN, USA
\and 
INAF-- Osservatorio Astronomico di Bologna, Via Ranzani, 1, 40127, Bologna, Italy
\and 
Dipartimento di Fisica e Astronomia, Via Ranzani, 1, 40127, Bologna, Italy
\and 
Nicolaus Copernicus Astronomical Center, Polish Academy of Sciences, Bartycka 18, 00-716, Warsaw, Poland
\and 
Massachusetts Institute of Technology, Kavli Institute of Astrophysics \& Space Research, Cambridge, MA, USA
\and
Section of Astrophysics, Astronomy and Mechanics, Department of Physics, University of Athens, 15784 Athens, Greece
\and
IAASARS, National Observatory of Athens, 15236 Penteli, Greece
\and 
INAF-- Osservatorio Astronomico di Padova, Vicolo dell'Osservatorio 5, 35122 Padova, Italy
\and 
INAF-- Osservatorio Astrofisico di Arcetri, Largo E. Fermi, 5, I-50125, Florence, Italy
\and 
Dipartimento di Fisica e Astronomia, Universita di Padova, vicolo Osservatorio 3, 35122 Padova, Italy
\and
Institute of Theoretical Physics and Astronomy, Vilnius University, A. Gostauto 12, 01108 Vilnius, Lithuania
\and
Centro de Astrobiolog\'{\i}a (INTA-CSIC), Departamento de Astrof\'{\i}sica, PO Box 78, E-28691, Villanueva de la Ca\~nada, Madrid, Spain
\and
Suffolk University, Madrid Campus, C/ Valle de la Vi\~na 3, 28003, Madrid, Spain
\and
INAF--Osservatorio Astrofisico di Catania, via S. Sofia 78, 95123 Catania, Italy
\and
Departamento de Astronom\'{i}a, Casilla 160-C, Universidad de Concepci\'{o}n, Concepci\'{o}n, Chile
\and
European Southern Observatory, Alonso de Cordova 3107 Vitacura, Santiago de Chile, Chile
\and
Centro de Estudios de F\'{i}sica del Cosmos de Arag\'{o}n (CEFCA), Plaza San Juan 1, E-44001 Teruel, Spain
\and
Institute of Astronomy, University of Cambridge, Madingley Road, Cambridge CB3 0HA, United Kingdom
\and
Astrophysics Group, Keele University, Keele, Staffordshire ST5 5BG, United Kingdom   
\and
INAF-- Osservatorio Astronomico di Palermo, Piazza del Parlamento 1, 90134, Palermo, Italy
\and
 ASI Science Data Center, Via del Politecnico SNC, 00133 Roma, Italy
 \and
Instituto de F\'isica y Astronomi\'ia, Universidad de Valparai\'iso, Chile
\and
Instituto de Astrof\'{i}sica de Andaluc\'{i}a-CSIC, Apdo. 3004, 18080 Granada, Spain
\and
N\'ucleo de Astronom\'ia, Facultad de Ingenier\'ia, Universidad Diego Portales,  Av. Ejercito 441, Santiago, Chile
\and
Departamento de Ciencias Fisicas, Universidad Andres Bello, Republica 220, Santiago, Chile
\and
Instituto de Astrof\'isica e Ci\^encias do Espa\c{c}o, Universidade do Porto, CAUP, Rua das Estrelas, 4150-762 Porto, Portugal}

\authorrunning{Overbeek et al.}
 
  \abstract
   {Trumpler 23 is a moderately populated, intermediate-age open cluster within the solar circle at a R$_{\mathrm{GC}} \sim$ 6 kpc. It is in a crowded field very close to the Galactic plane and the color-magnitude diagram shows significant field contamination and possible differential reddening; it is a relatively understudied cluster for these reasons, but its location makes it a key object for determining Galactic abundance distributions.}
   {New data from the Gaia-ESO Survey enable the first ever radial velocity and spectroscopic metallicity measurements for this cluster. We aim to use velocities to isolate cluster members, providing more leverage for determining cluster parameters.}
   {Gaia-ESO Survey data for 167 potential members have yielded radial velocity measurements, which were used to determine the systemic velocity of the cluster and membership of individual stars. Atmospheric parameters were also used as a check on membership when available. Literature photometry was used to re-determine cluster parameters based on radial velocity member stars only; theoretical isochrones are fit in the $V$, $V-I$ diagram. Cluster abundance measurements of ten radial-velocity member stars with high-resolution spectroscopy are presented for 24 elements. These abundances have been compared to local disk stars, and where possible placed within the context of literature gradient studies.}
   {We find Trumpler 23 to have an age of 0.80 $\pm$ 0.10 Gyr, significant differential reddening with an estimated mean cluster E($V-I$) of 1.02$^{+0.14}_{-0.09}$, and an apparent distance modulus of 14.15 $\pm$ 0.20. We find an average cluster metallicity of [Fe/H] = 0.14 $\pm$ 0.03 dex, a solar [$\alpha$/Fe] abundance, and notably subsolar [s-process/Fe] abundances.}
   {}
\maketitle

\section{Introduction} \label{intro}

Open clusters (OCs) are very useful tools for the study of physics on stellar and Galactic scales. The astrophysical properties of stars can be isolated by studying a single-age population of stars in multiple evolutionary states; nucleosynthesis can also be probed with precise measurements of abundances of clusters in different age ranges. Because the dispersion in abundances between unevolved stars in a single cluster is typically small or nonexistent \citep[see, e.g.,][]{Randich06, deSilva06, Liu16}, many cluster stars can be used to gain a more accurate measurement of the primordial cluster abundance. Since a cluster is composed of a single age population, the evolutionary states of stars can be relatively easily determined with a color-magnitude diagram. Cluster parameters (age, distance, reddening, metallicity) can also be determined in this way, allowing clusters to be placed within the larger context of the evolution of the Galaxy.
\\
\indent A current topic of active research is the determination of radial abundance gradients for various groups of chemical elements across different regions of the Galactic disk. Open clusters and field stars can both be used as tracers of the chemical history of the disk, but clusters have more easily determined ages and distances, putting constraints on different epochs of Galaxy formation. They are also found at a large range of Galactocentric radii and ages, ranging from millions of years old to approximately 10 Gyr old \citep{Salaris04} and $\sim$5 to 24 kpc from the center of the Galaxy \citep{Magrini10, CC07}. Recent efforts have been made to provide a uniform age scale for OCs \citep{Salaris04, BT06}, and to probe the outer regions of the disk \citep[e.g.,][]{Carraro04, Bragalia08, Sestito08, Yong12, CantatGaudin16}, the solar neighborhood \citep[e.g.,][]{Frinchaboy13}, and the inner disk \citep{Magrini10, Jacobson16}. However, relatively few OCs well inside the solar circle (R$_{\mathrm{GC}}$ $<$ 7 kpc) have parameter determinations due to observational challenges, and the measurements that do exist can be difficult to compare from study to study.
\\
\indent Several large studies address the need for observational data of many clusters exploring the Galactic parameter space on a common analysis and abundance scale, such as RAVE \citep{Conrad14}, LAMOST \citep{Deng12}, OCCASO \citep{Casamiquela16}, and OCCAM as part of APOGEE \citep{Frinchaboy13, Cunha16}. The Gaia-ESO Survey (GES) is a public spectroscopic survey of stars in all components of the Milky Way using VLT FLAMES planned as a complement to the Gaia mission \citep{Gilmore12, Randich13}. Observations and data analysis are currently underway, and the completed survey will cover some 100,000 stars in the Galaxy and $\sim$70 OCs. The GES will measure at least 12 elements (Na, Mg, Si, Ca, Ti, V, Cr, Mn, Fe, Co, Sr, Zr, and Ba) for a few thousand field stars and the full sample of OCs with high-resolution spectroscopy. This is one of a series of papers on OC data obtained and analyzed by GES focusing on the intermediate-age inner disk cluster Trumpler 23.
\\
\indent Trumpler 23 is a moderately well-populated OC located in the 4th quadrant of the Galaxy at $l$ = 328.86$^{\circ}$, $b$ = -0.47$^{\circ}$, close to the plane of the Galaxy toward the Galactic center. It is approximately 0.6 Gyr old and 2 kpc from the Sun \citep{Carraro06}, well inside the solar circle at an R$_{\mathrm{GC}}$ $\sim$ 6 kpc. Due to previously noted complications with crowding, differential reddening (DR) and possible tidal effects \citep{BB07} membership determination via photometry is difficult and no previous spectroscopic studies exist. However, its location and characteristics make it an interesting candidate for further study. The GES is well positioned to take full advantage of Tr 23 observations because of its uniform data analysis pipelines and large potential number of comparison clusters at various Galactocentric radii \citep[Jacobson et al. 2016]{Magrini14}. 
\\
\indent The paper is structured as follows: in Section \ref{lit_data} we review previous literature studies of Trumpler 23; in Section \ref{targets} we describe the structure of the Gaia-ESO Survey and observations; in Section \ref{members} we perform radial velocity membership selection; in Section \ref{atm_params} we detail atmospheric parameter determinations of target stars; in Section \ref{cl_params} we determine our own set of cluster parameters using literature photometry; in Section \ref{abuns} we discuss abundance measurements; and in Section \ref{summary} we summarize our points and conclude.

\section{Cluster parameters from the literature} \label{lit_data}

\citet{Trumpler30} first described Tr 23 as a sparse cluster of relatively faint stars not easily distinguished from its surroundings. The first data on Tr 23 were provided by \citet{vdBH75} as part of a two-color survey of southern OCs; they classified it as a moderately well-populated cluster with a diameter of 6$\arcmin$.
\\
\indent There are two existing photometric studies of Trumpler 23 by \citet{Carraro06} and \citet{BB07}. Carraro et al. obtained Johnson-Cousins $V, I_c$ photometry for $\sim$11000 stars in the field of Tr 23. They noted that the main sequence is unusually broad, which they attributed to DR or perhaps a high cluster binary fraction. They also pointed out that Tr 23 may be undergoing strong tidal interactions with the Galaxy because of its apparent elongated shape on the sky. They used Padova isochrones \citep{Girardi00_iso} to estimate a cluster age of 0.6 $\pm$ 0.1 Gyr assuming solar metallicity, although they noted that the age could vary from 0.5 to 1.5 Gyr depending on the true cluster metallicity. They also found a high reddening E($V-I$) = 1.05 $\pm$ 0.05 (converted to an E($B-V$) = 0.84), and derived cluster parameters of (m $-$ M)$_{V}$ = 14.35 $\pm$ 0.20, and d$_{\odot}$ = 2.2 kpc (corresponding to a Galactocentric radius of 6.2 kpc if R$_{\odot}$ = 8.0 kpc). 
\\
\indent \citet{BB07} examined six clusters inside the solar circle that are located in crowded fields, including Tr 23. They used 2MASS $J$, $H$, and $K_s$ photometry to correct for field contamination and derive cluster parameters for Tr 23. In the cleaned CMD, the Tr 23 main sequence in $J$, $J-H$ is still unusually broad. The authors found evidence of DR in an 80$\arcmin$ $\times$ 80$\arcmin$ field around Tr 23. The profile of the cluster after color-magnitude filtering is irregular across declination, and field star counts drop significantly toward higher declinations. The radial distribution of assumed photometric members in the Tr 23 field is not well-fit by a two-parameter King model, and does not appear to be a smooth distribution, which they also attribute to significant DR. The authors also noted that the inner disk clusters studied have unusually high mass densities and small limiting radii compared to solar neighborhood OCs, which would suggest strong Galactic tidal effects on Tr 23. Using Padova isochrones of solar metallicity as in \citet{Carraro06}, they found a significantly lower reddening of E($B-V$) = 0.58 $\pm$ 0.03, an age of 0.9 $\pm$ 0.1 Gyr, and a distance d$_{\odot}$ = 1.9 $\pm$ 0.1 kpc.

\section{Observations} \label{targets}
\subsection{Gaia-ESO survey methods}

Data for GES are taken with FLAMES (Fiber Large Array Multi-Element Spectrograph) on the VLT at the European Southern Observatory. FLAMES has two instruments, the medium-resolution multi-object spectrograph GIRAFFE and the high-resolution UVES (Ultraviolet and Visual Echelle Spectrograph) \citep{Pasquini02}. The target selection, observation, data reduction, atmospheric parameter determination, and abundance measurements are handled by specific working groups (WGs) within the collaboration. Parameter and abundance determinations for each target are typically done by multiple subgroups within WGs called abundance analysis nodes, and the results of individual nodes are combined within each WG; the WG values are then homogenized to yield final recommended parameters. This structure produces homogenous parameter determinations while allowing WGs to specialize in different types of stars. The data described here come from the fourth internal data release (GESviDR4Final) which comprises a (re-)analysis of all available spectra taken before July 2014 using an updated linelist \citep{Heiter15} and analysis techniques. For more details about the data reduction, which will not be discussed here, see \citet{Sacco14} and Lewis et al. (in prep).
\\
\indent Analysis of the GIRAFFE FGK star atmospheric parameters and abundances \citep[using solar abundances by][]{Grevesse07} is handled by WG 10 (Recio-Blanco et al., in prep), UVES FGK star parameters and abundances by WG 11 \citep{Smiljanic14}, and the homogenization of results from different subgroups by WG 15 (Hourihane et al., in prep). \citet{Smiljanic14} describes in detail the methods used by individual nodes and the process of obtaining stellar parameters and abundances for FGK stars using UVES data. Multiple WG11 nodes measure abundances and atmospheric parameters using the same atomic data, solar abundances, and model atmospheres, but different methods. Some nodes determine atmospheric parameters using equivalent widths and minimizing abundance trends with excitation potential, line strength, etc., and some nodes compare observed spectra to spectral libraries. For abundance determinations, some nodes use automated or semi-automated equivalent width determinations and some use spectral synthesis software. The abundance homogenization process of GES makes use of a number of calibrating objects (clusters with many stars) to assess the precision of each abundance analysis technique, and Gaia benchmark stars (individually well-studied stars with independently determined atmospheric parameters) to test the accuracy of abundance measurements and atmospheric parameters (Pancino et al., in prep.). Node measurements of benchmark and calibration stars are used to evaluate the performance of each node in different parameter spaces; a final WG value for each star is determined by weighting each node's value by its accuracy in determining benchmark parameters.  These weights are also used in determining final abundances. The evaluation of calibration and benchmark star abundances determined by individual nodes also ensure that uncertainties given for stellar abundances reflect the inherent uncertainties in the various measurement techniques.

\subsection{Target selection for GES}

\begin{figure*}
\centering
\includegraphics[width=0.7\hsize]{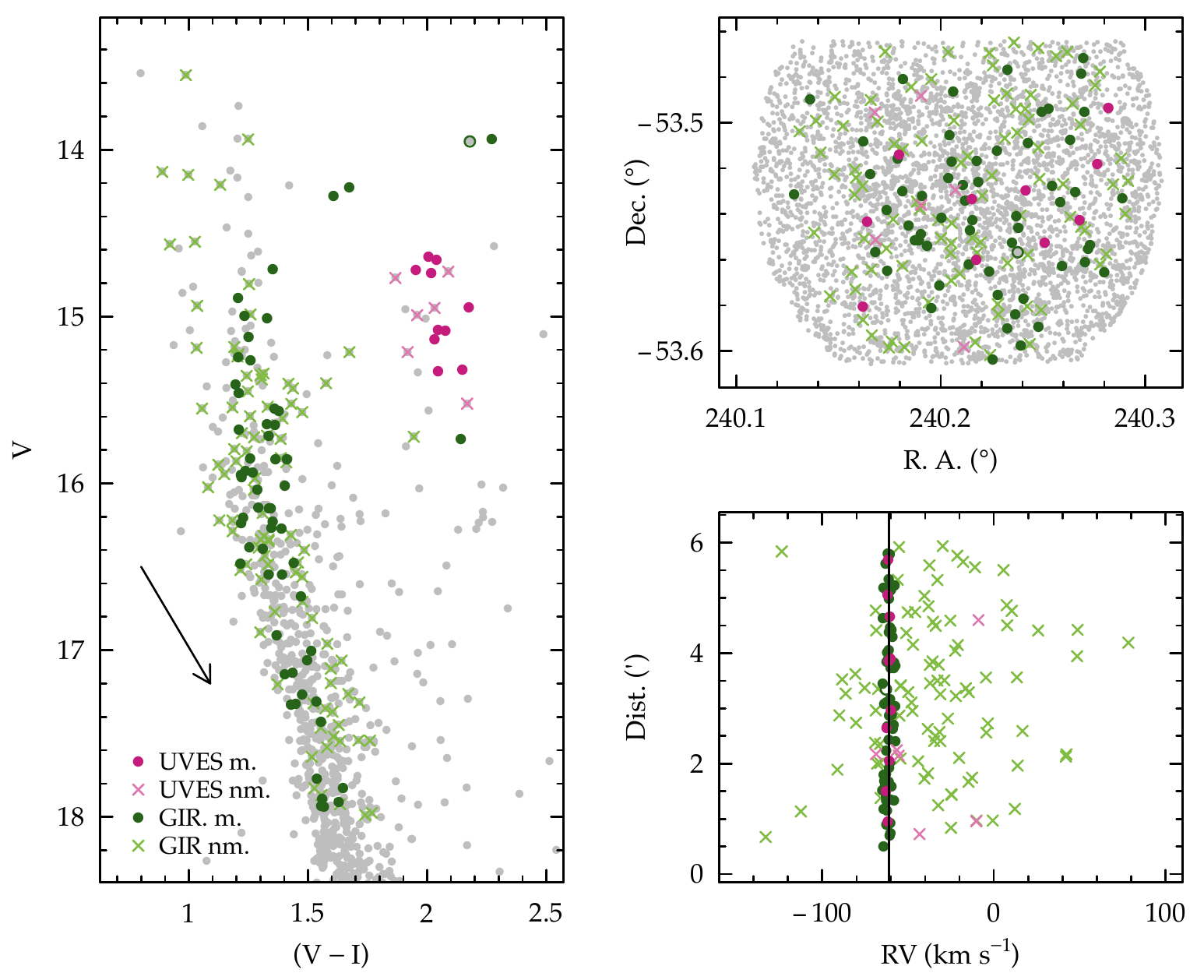}
\caption{\textit{Left:} CMD of Trumpler 23 of photometry from \citet{Carraro06} with all stars within 6$\arcmin$ marked as small grey dots, UVES cluster members as purple circles, UVES non-members as light purple crosses, GIRAFFE cluster members as dark green circles, and GIRAFFE non-members as light green crosses. Star 18, targeted by GIRAFFE, is a cluster member by radial velocity but has inconsistent atmospheric parameters (see Section \ref{cl_params}) so it is marked with an open green circle. A reddening vector is also marked in the lower left. \textit{Top right:} the same symbols showing the coordinates of all stars observed in the Gaia-ESO survey and  \citet{Carraro06} stars within 6$\arcmin$. \textit{Lower right:} the same symbols showing distance from the cluster center vs. radial velocity of all observed stars except for one GIRAFFE non-member at +459 \kms. The cluster systemic velocity is marked as a solid line.}
\label{CMD}
\end{figure*}

\indent For intermediate-age OCs with prominent red clumps, GES targets are selected as follows: likely clump stars are observed with UVES, so that the most time-intensive targets are most likely to be members, followed by probable red giants if the clump is sparse. Main sequence stars down to $V$ = 19 are observed with GIRAFFE, using the HR9B setup primarily for stars of spectral type A to F and the HR15N setup for cooler stars. General GES target selection methods are outlined in Bragaglia et al. (in prep.).
\\
\indent The field of Tr 23 is crowded, reaching 50\% field star contamination at 5$\arcmin$ \citep{Carraro06}; careful target selection is required. Targets were chosen based on both the \citet{Carraro06} $VI$ photometry and 2MASS $J$, $H$, and $K$. 151 apparent main-sequence and giant stars were selected as GIRAFFE targets with 13.5 $<$ $V$ $<$ 18 within 6$\arcmin$ of the cluster center in order to limit field contamination, which increases with both distance from the cluster center and $V$ magnitude. Tr 23 has a distinct red clump with relatively high membership probabilities, so 16 stars out of 21 in or near the clump and within 6$\arcmin$ of the cluster center were selected as UVES targets. 
\\
\indent Data were taken on the nights of 28 and 29 July and 13 and 14 September 2013. 54 GIRAFFE targets were observed with the HR15N setup only (6470-6790\A~and R = 17,000), 81 stars were observed with the HR9B setup only (5143-5356\A~and R = 25,900), and 16 targets were observed with both. UVES targets were observed with the U580 setup (4800-5800\A~and R = 47,000). The typical signal-to-noise ratios achieved for the HR15N, HR9B, and U580 setups are $\sim$160, 60, and 90, respectively. The CMD and coordinates of target stars are shown in Figure \ref{CMD}.

\section{Membership determination} \label{members}

\begin{figure*}
\centering
\includegraphics[width=0.8\hsize]{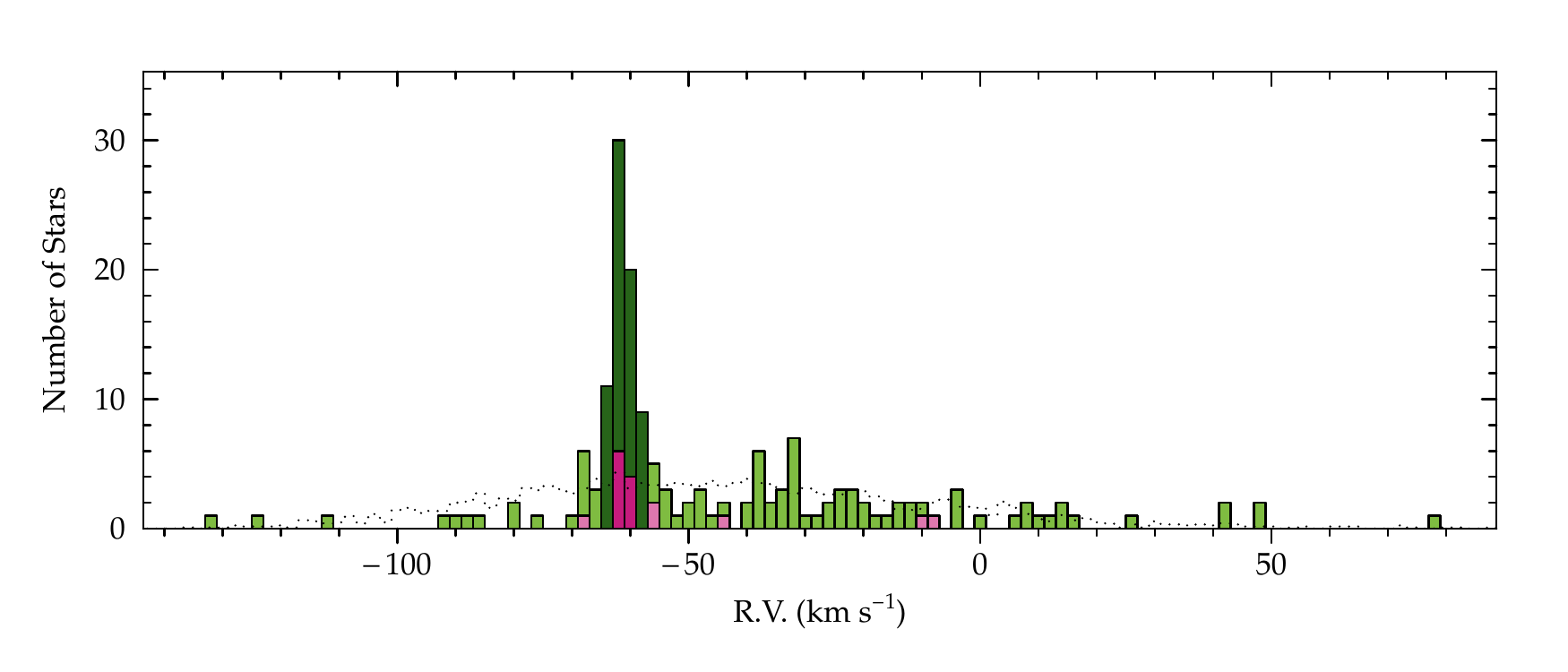}
\caption{Stellar radial velocity distribution of GIRAFFE and UVES target objects using the same color scheme as in Figure \ref{CMD}. The dotted line represents a Besan\c{c}on star count model for the same field as the Carraro et al. (2006) photometry, scaled as the number of field stars observed in GES divided by the total number of field stars in the same color and magnitude range.}
\label{hist}
\end{figure*}

Because Tr 23 has heavy field star contamination, radial velocities are very important in determining membership. GIRAFFE radial velocities have a typical error of $\sim$0.4 \kms, although stars with the highest rotational velocities or lowest SNR have errors up to several \kms~\citep{Jackson15}; UVES radial velocity errors are around 0.6 \kms~ \citep{Sacco14}. Figure \ref{hist} shows a histogram of UVES and GIRAFFE target velocities from $-140$ to 80 \kms~(excluding the most discrepant GIRAFFE star). There is a clear peak in the velocity distribution around $-60$ \kms, but a wide range of field star velocities. Using an iterative 2$\sigma$ clipping technique on the mean until $< 5\%$ of stars were eliminated as outliers (13 iterations), we have determined a cluster velocity of $-61.3 \pm 1.9$ \kms~(s.d.) from 70 member stars within 2$\sigma$ of the mean. Thus, ten out of the total of 16 UVES target stars have radial velocities consistent with the cluster compared to 60 out of 151 GIRAFFE targets (20 observed with HR15N setup only, 21 observed with HR9B setup only, and nine observed with both). The cluster radial velocity dispersion of 1.9 \kms~ is slightly higher than typical for OCs; \citet{Mermilliod09} find a typical dispersion on cluster radial velocities of $\sim$1 \kms. However, considering the errors on the velocities and possible contamination by field stars and unresolved binaries, this is a reasonable dispersion.
\\
\indent As is clear from Fig. \ref{hist}, the field of Tr 23 has a significant number of field stars at a range of radial velocities, some of which will fall within the cluster radial velocity range (the minimum radial velocity of member stars selected by sigma clipping is $-64.9$ \kms~and the maximum is $-57.3$ \kms). We have used a Besan\c{c}on star count model (http://model.obs-besancon.fr/) to estimate the number of field stars falling within the cluster radial velocity limits \citep{Robin03}. Figure \ref{hist} shows the observed radial velocity distribution overlaid with the field radial velocity distribution predicted by the model. The predicted distribution is scaled to the total number of stars observed divided by the number of stars in the \citet{Carraro06} photometry in the same $V$ mag range. The scaled model predicts that $\sim$15 field stars fall within the selected bounds of the cluster radial velocity distribution. This is a rough scaling, but it reproduces reasonably well the field star velocity distribution, including the long tail of the observed field star distribution toward higher radial velocities typically observed in the disk. The reverse scenario, where member stars which are spectroscopic binaries happen to fall outside of the cluster radial velocity distribution, is also possible; however, this possibility will not affect conclusions drawn in this work.

\section{Atmospheric parameters} \label{atm_params}

\begin{figure*}
\centering
\includegraphics[width=0.4\hsize]{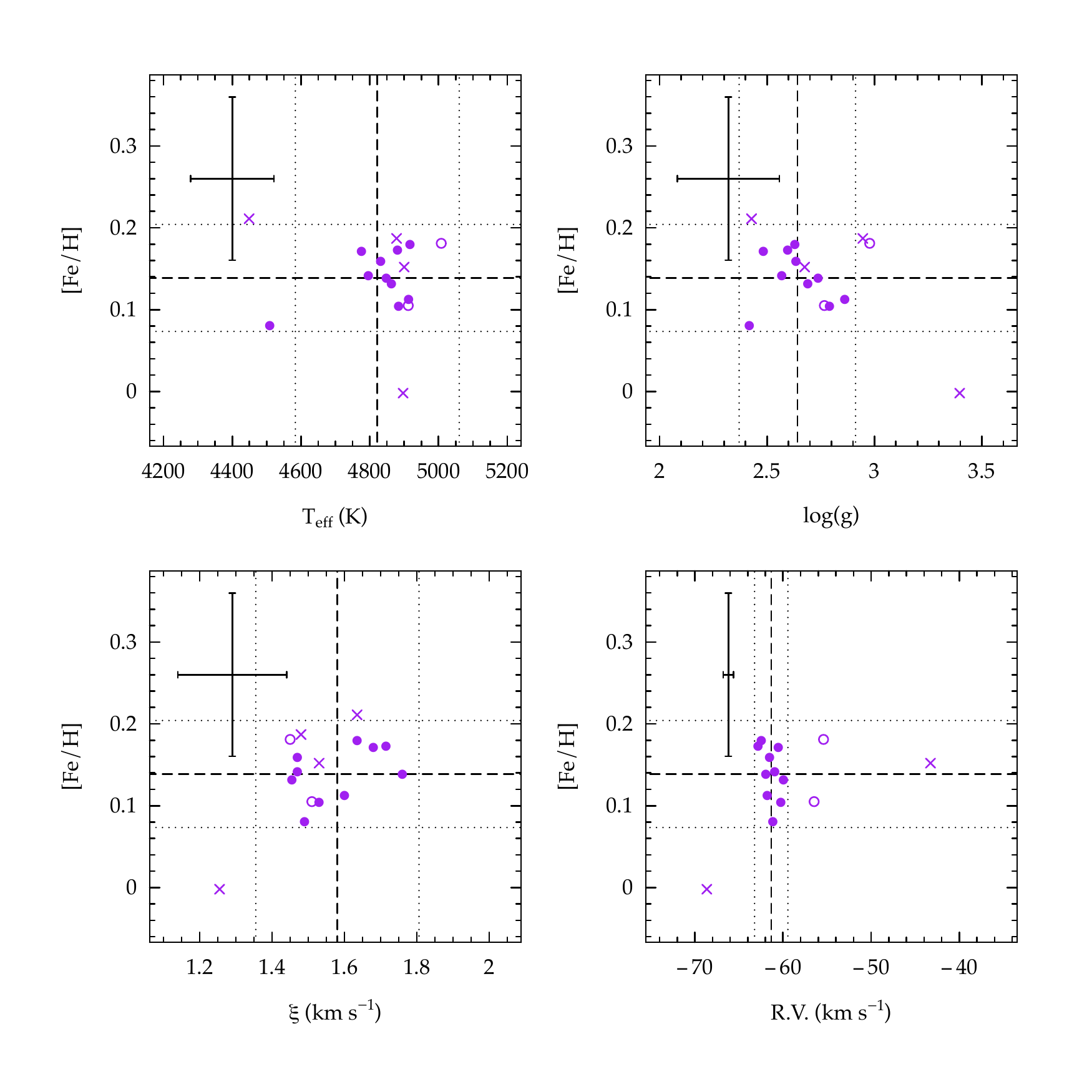}
\caption{UVES target [Fe/H] against radial velocity. Ten RV members are marked as filled circles, two possible binary members as open circles, and two non-members as crosses (two additional non-members lie outside the bounds of this plot). Dashed and dotted lines indicate the mean and 2$\sigma$ boundaries respectively.}
\label{params}
\end{figure*}

Effective temperatures were determined for 31 of the 60 GIRAFFE radial velocity members; atmospheric parameters (when available) are listed along with metallicities, coordinates, photometry, and velocities in Table \ref{gir_info} (here and in other tables and figures we use $\xi$ to represent microturbulent velocity). Due to methods of pipeline analysis and limitations of the data, some GIRAFFE stars do not have atmospheric parameter determinations, or only partial parameter sets. 46 out of 91 GIRAFFE radial velocity non-members also have at least partial parameter determinations; these are given in Table \ref{gir_nm_info}.
\\
\indent Figure \ref{params} shows UVES [Fe/H] plotted against radial velocities for 14 UVES targets with fully determined parameters; two UVES targets with highly discrepant radial velocities are not shown. Radial velocity (RV) members are marked as filled circles, potential binary members as open circles, and non-members as crosses. Dashed lines indicate the UVES member mean values and dotted lines mark 2$\sigma$ from the mean.
\\
\indent Star 74 has a radial velocity 3$\sigma$ from the determined cluster velocity, and star 86 is 2.5$\sigma$ from the cluster radial velocity. Stars 74 and 86 may be binary members, but we will consider them non-members for the purpose of evaluating cluster parameters. Stars 56, 89, 131, and 147 are more than 4$\sigma$ from the mean cluster radial velocity, so we consider them non-members. Table \ref{uves_info} gives UVES target photometry, coordinates, radial velocities, atmospheric parameters, and membership status.
\\
\indent The weighted mean [Fe/H] of the ten UVES radial velocity cluster members is 0.14 $\pm$ 0.03 (s.d.), with a typical stellar [Fe/H] error of 0.10 dex. The weighted mean of the 20 GIRAFFE radial velocity cluster members with abundance determinations is 0.15 $\pm$ 0.17 (s.d.) with a typical stellar [Fe/H] error of 0.28 dex. The two determinations show excellent agreement, suggesting that the sometimes systematic differences in UVES and GIRAFFE abundances seen in earlier releases have been eliminated. The weighted average of both the UVES and GIRAFFE radial velocity member [Fe/H] is 0.14, with an error on the mean of 0.03 dex; we have adopted this as the cluster [Fe/H]. 
\\
\indent With a Galactocentric distance of only 6 kpc, Tr 23 is one of the few intermediate-age clusters we can use to probe the inner Galactic disk. Its clear super-solar metallicity makes it especially interesting in the context of the behavior of the abundance gradient inside the solar circle. \citet{Magrini10}, from a new sample of three inner disk open clusters, along with a selection of clusters from the literature, found evidence that the gradient increased strongly toward the Galactic center, but their sample included only one cluster inside 6.5 kpc, and only three inside 7 kpc. The GES survey now provides abundances on a uniform metallicity scale for nine clusters inside this limit, with Tr 23 being among the three closest to the Galactic center. As shown in \citet{Jacobson16}, this GES cluster sample can be described by a linear relationship of increasing metallicity with decreasing Galactocentric distance with a slope of -0.10 $\pm$0.02 dex kpc$^{-1}$. The abundance of Tr 23 falls neatly on this relationship, substantiating the gradual rather than a steepening increase in metallicity in these innermost regions.  

\section{Re-determined cluster parameters} \label{cl_params}

\begin{figure*}
\centering
\includegraphics[width=0.7\hsize]{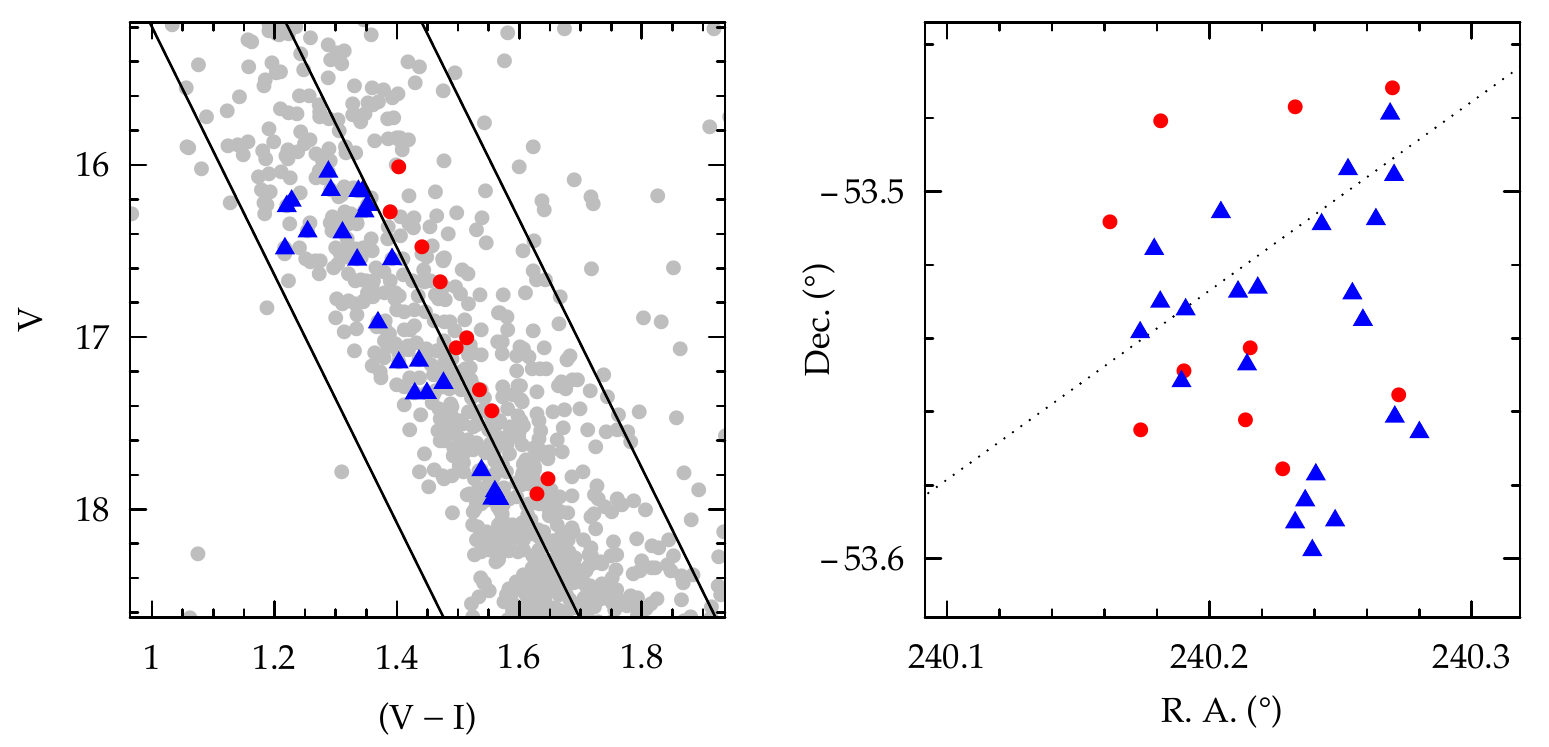}
\caption{\textit{Left:} a zoomed-in version of the CMD in Figure \ref{CMD}, with possible blue and red main sequences marked as triangles and circles. \textit{Right:} plot of the right ascension and declination of these stars using the same symbols for each group, with a dotted line marking constant Galactic latitude b = -0.47.}
\label{diff_red}
\end{figure*}

With radial velocity membership and [Fe/H] determinations as a guide, we have determined a refined set of cluster parameters from the member CMD. Member selection narrows the main sequence (MS) somewhat (see Fig. \ref{CMD}), although the width still allows for some degeneracy in CMD fitting. \citet{Carraro06} suggest that stars in the field of Tr 23 past the main-sequence turn-off (TO) might be blue stragglers, but all of the objects observed by GES in this region are radial velocity non-members, so these may in fact be mostly field stars. None of the seven potential blue stragglers observed have determined metallicities; only one has a RV within 4$\sigma$ of the cluster average, and the others are more than 10$\sigma$ from the cluster RV. There are also four GIRAFFE radial velocity members that appear to be evolved stars based on their photometry. Two of these stars have determined atmospheric parameters: one appears to be on the subgiant branch and has parameters that are consistent with its location on the CMD. The other, star 18, which is marked as an open circle in Figure \ref{CMD}, appears to be on either the giant branch or AGB but has parameters that would indicate it is actually a dwarf; it is likely an interloper hidden in the radial velocity distribution of the cluster. We did not include this star in the isochrone fitting. The two apparent giants without atmospheric parameter determinations were treated as members.
\\
\indent The red clump is pronounced and well-populated, but also appears unusually extended, covering $\sim$0.7 magnitudes in $V$ and 0.3 in $V-I$. As mentioned above, the width of the MS ($\sim$0.2 in $V-I$) is much larger than the photometric errors for $V$ $<$ 18, and may be attributable to DR or a high cluster binary fraction. We discuss the binary possibility at the end of the section. If the cause of the width of the MS is DR, one might expect stars on the redder side of the MS to appear in different locations on the sky than stars on the apparent blue side of the MS. Figure \ref{diff_red} shows main sequence members with 16 $<$ $V$ $<$ 18 divided into two groups, a blue edge and red edge of the MS, and the location of stars in these two groups on the sky. The two color groups appear to separate into four stripes on the sky, with the DR mainly occurring along the declination axis as found by \citet{BB07}, although the size of the cluster (and the plot in Figure \ref{diff_red}) is significantly smaller than the $80\arcmin \times 80\arcmin$ area on the sky they use to examine the DR. The Galactic plane is at an angle of $\sim$ -27$^{\mathrm{o}}$ from the axis of right ascension in this figure, so the direction of the DR corresponds to Galactic latitude, with stars on the red edge of the MS appearing in two groups at higher average latitudes than stars on the blue edge of the MS. At a distance of 2.2 kpc the angular extent of radial velocity members corresponds to 6 parsecs. Tr 23 is only 50 pc above the plane, so it is likely that reddening would vary with shallow angles out of the plane; DR of this magnitude (A$_{V} \sim 0.5$) in the direction of the bulge has been reported across small fields on the order of parsecs \citep{Ortolani90, Gonzalez12}. The spread in $V$ and $V-I$ of the clump also roughly correspond with the dimensions of the reddening vector shown in Figure \ref{CMD} (left panel).

\begin{figure*}
\centering
\includegraphics{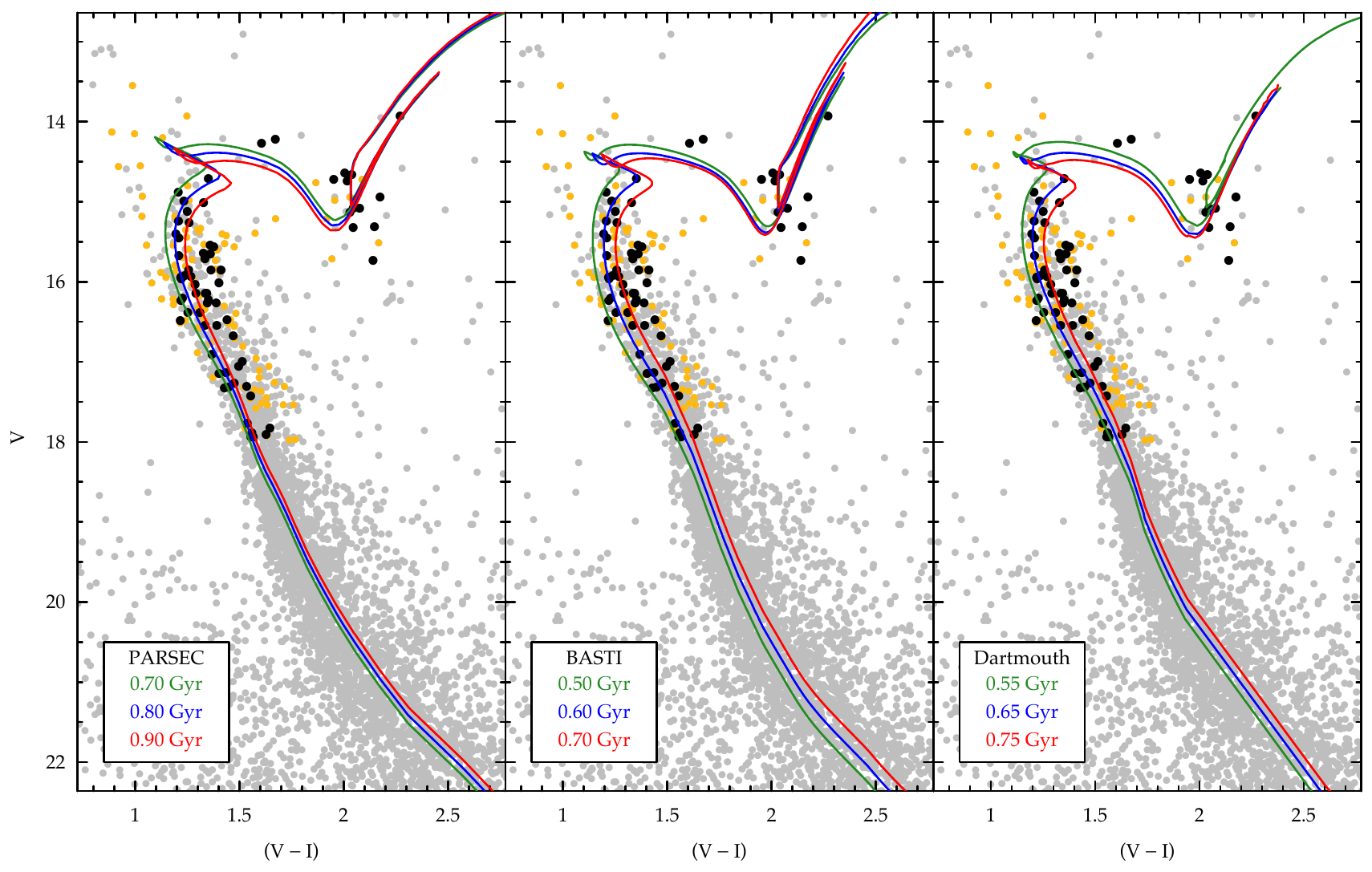}
\caption{Tr 23 CMD with Carraro et al. (2006) photometry in grey, cluster members in black circles, non-members in orange dots, and isochrones. Left panel is PARSEC isochrones of 0.7, 0.8, and 0.9 Gyr; middle panel is BASTI isochrones of 0.5, 0.6, and 0.7 Gyr; right panel is Dartmouth isochrones of 0.55, 0.65, and 0.75 Gyr in green, blue, and red, respectively. Adopted reddening and distance moduli are given in Table \ref{iso_params}.}
\label{iso_mag}
\end{figure*}

\indent We have attempted to correct for DR in the field of Tr 23. For a complete discussion of the methods used, we refer the reader to \citet{Donati14}; briefly, we selected an area of the MS below the TO but above areas of highest field contamination and set a fiducial line along the MS, then iteratively calculated the median DR for each MS star and a group of its nearest neighbors, and binned the results on the sky to get the final corrections. However, our corrections for the DR effects did not yield a marked improvement in the definition of the MS when considering all stars in the field or the member stars only (which go out to 6$\arcmin$ from the cluster center; see Fig. \ref{CMD}). Therefore, we fitted isochrones and derived cluster parameters based on the uncorrected photometry, fitting the blue edge of the MS which seems reasonably well defined. We do not have membership determinations for enough clump stars to look at the possibility of DR as a cause of its extended $V$ mag range, but seven of the ten apparent clump stars fall in a fairly narrow range in $V-I$ of 2.00 to 2.08. We can thus use the red clump for a reasonable assessment of the general reddening across Tr 23, though we estimate the error in the reddening conservatively. The algorithm we use for the DR correction does consistently estimate a reddening variability of $\pm$0.1 in $V-I$, which matches well the width of the clump in $V-I$ and our estimated uncertainty on our E($V-I$) discussed below.
\\
\indent We have used PARSEC \citep{Bressan12}, BaSTI \citep{Pietr04}, and Dartmouth \citep{Dotter08} isochrones to determine the age, reddening, and distance of Tr 23 based on the selected member stars and cluster metallicity. Metallicity values were set relative to a solar Z = 0.0152 where possible; an input of Z = 0.021 was used for the PARSEC and Dartmouth models, and the closest option of Z = 0.0198 was selected for the BaSTI models. The isochrones were also selected to have a solar [$\alpha$/Fe] ratio (see Section \ref{alpha}). Figure \ref{iso_mag} shows the cluster CMD with \citet{Carraro06} photometry for stars within 6$\arcmin$ of the cluster center in gray, cluster members as black circles, non-members as orange dots, and selected isochrones for each model set in blue, with intervals corresponding to $\pm$0.1 Gyr in red and green. Table \ref{iso_params} gives the best fit parameters for each isochrone, and values found in the literature. 

\begin{figure*}
\centering
\includegraphics{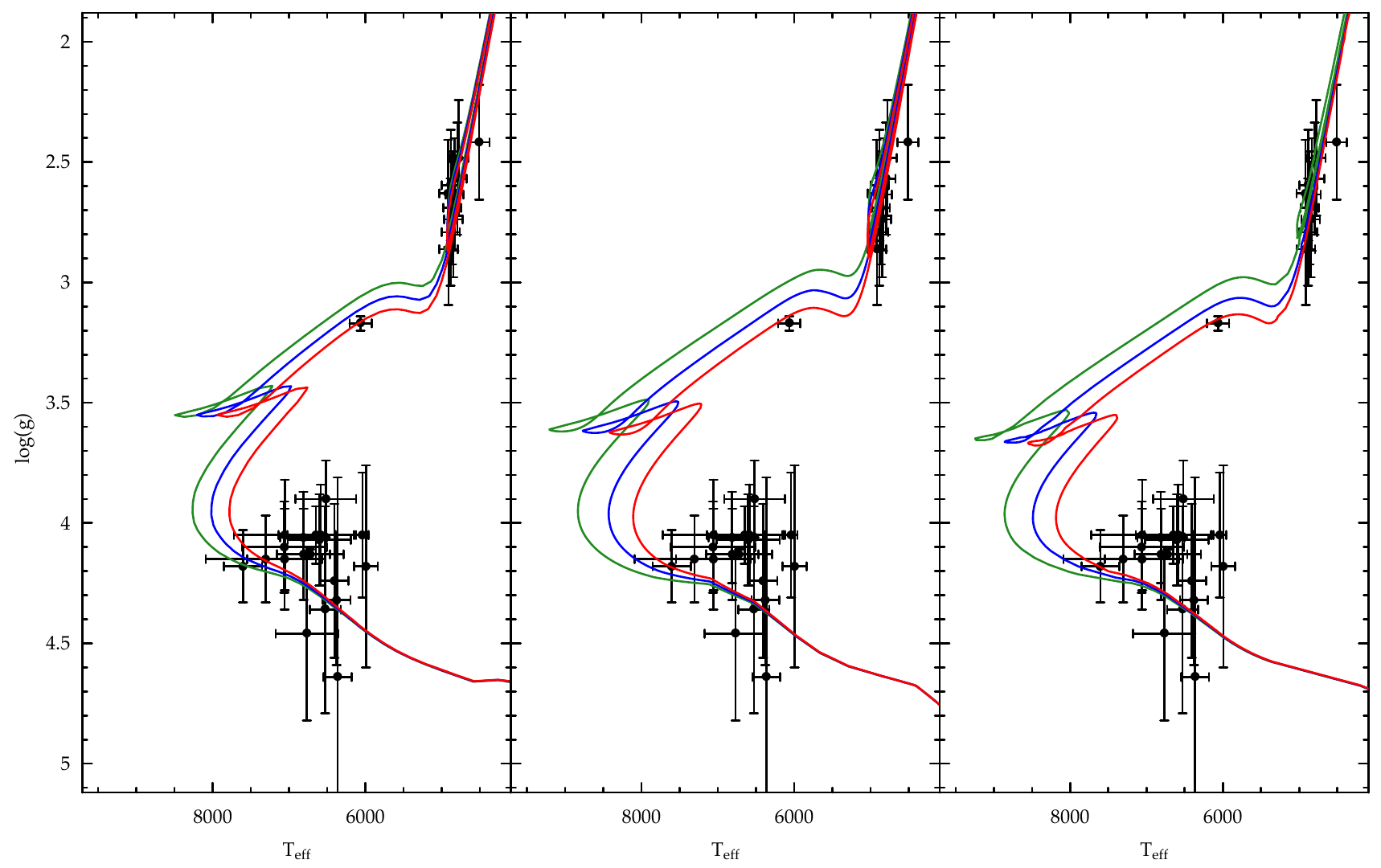}
\caption{Tr 23 gravities vs. effective temperatures of cluster members in black with isochrones. Panels and lines the are same as in Figure \ref{iso_mag}.}
\label{iso_par}
\end{figure*}

\indent The PARSEC 0.8 Gyr isochrone provides a good fit to the TO and a reasonable fit to the blue edge of the lower MS. The reddening can be adjusted to provide a good fit to the center of the red clump, as well as the star that appears to be on the red giant branch. The E($V-I$) for this fit is 1.02$^{+0.14}_{-0.09}$, where errors are based on fitting either edge of the clump as defined by the ten UVES radial velocity members; two of these that are separated from the group in ($V-I$) cannot be well fit by any of the isochrones while also fitting the main clump. This is likely due to DR, so the errors on our reddening estimates encompass the DR effects on the field. We have used the extinction coefficients of \citet{Dean78} to derive an E($B-V$) of 0.82$^{+0.11}_{-0.08}$. The apparent distance modulus for this fit is (m-M)$_V$ = 14.15, with a corresponding solar distance of 2.1 kpc.
\\
\indent The BaSTI 0.6 Gyr isochrone provides a good fit to the lower main sequence and TO region, but the giant and subgiant branches are not as well fit. Here, we use the BaSTI models with convective overshooting in order to obtain a good fit to the lower main sequence. The E($V-I$) required to match this isochrone to the red clump is 1.09$^{+0.14}_{-0.09}$ with a corresponding E($B-V$) of 0.87$^{+0.11}_{-0.08}$. The best fit distance modulus is 14.50 (d$_{\odot}$ = 2.3 kpc).
\\
\indent The Dartmouth 0.65 Gyr isochrone fits the turnoff region well, but the blue side of the lower main sequence is not well reproduced, and the red clump and AGB do not fit some stars. The 0.65 Gyr isochrone results in a E($V-I$) of 1.10$^{+0.14}_{-0.09}$ and an E($B-V$) of 0.88$^{+0.11}_{-0.08}$. The best fit distance modulus is (m-M)$_V$ = 14.45 (d$_{\odot}$ = 2.2 kpc). The distance moduli and solar distances we have derived from each of the three models are thus also in agreement.
\\
\indent We can check the ages we have determined with each set of isochrones by plotting HR diagrams. Figure \ref{iso_par} shows the same models as in Figure \ref{iso_mag} in log(g) vs. T$_{\mathrm{eff}}$ space, with GIRAFFE and UVES stars with atmospheric parameter determinations overplotted. The errors on many of the points are large enough that these diagrams are not extremely useful in determining the age, but all stars are consistent with these sets of parameters within their error bars and the uncertainties on the isochrone age. The PARSEC isochrones fit the clump star gravities better than the BaSTI and Dartmouth models, which appear to slightly underestimate clump gravities by $\sim$0.1 dex. We can also check our reddening determination with UVES effective temperatures; using the color-temperature relations of \citet{Alonso99}, we find $V-I$ colors for UVES member stars and use these to find a simple reddening estimate. In this way we find a typical E($V-I$) of 1.09 $\pm$ 0.07 (s.d.). This is consistent with the reddening derived by isochrone fitting for all three sets of isochrones, with a range in calculated E$(V-I)$ of 0.19. Errors in effective temperature of $\sim$100 K could also change the calculated E($V-I$) by 0.04 mag.

\indent It is notable that, although the UVES member stars occupy a small range in temperature ($\sim$150K excluding a single outlier in Fig. \ref{params}), there is a spread in log(g) of about 0.4 dex. The typical errors on stellar gravities for the UVES stars are $\sim$0.2 dex, so this is not a highly significant spread, but it is also worth considering the evolutionary effects that might cause a real spread in red clump gravities. \citet{Girardi00} examine the $V$ magnitude spread in clump stars in OCs NGC 752 and NGC 7789, and postulate that although OCs show no evidence of different age stellar populations that would lead to a stellar mass difference large enough to cause a separation in the clump, mass loss on the RGB is not well understood and it may be possible that some stars in an intermediate age OC would lose enough mass to undergo a helium flash while other cluster stars with slightly more mass would not. Stars that do experience a helium flash would of course be significantly more luminous than those that do not, and this effect would cause a difference in the $V$ magnitude of the lower mass stars as well. The limit for core helium flash to occur is $\sim$2.0 M$_{\odot}$, and our model isochrones predict a clump star mass for Tr 23 of 2.3 to 2.5M$_{\odot}$ without significant mass loss. This effect of a secondary red clump has only been observed in clusters older than 1 Gyr, most recently in NGC 419 \citep[$\sim$1.4 Gyr; see][]{Girardi09}, so this is a less likely explanation for the $V$ mag dispersion of the clump than DR or binary stars. There are several evolutionary effects that would cause mass loss prior to the red clump stage \citep[see][]{Donati14}, but these scenarios still would not involve mass loss of the degree necessary for an OC $<$ 1 Gyr old.
\\
\indent For all three models we estimate an error on the ages of 0.1 Gyr, as that is the smallest difference that visibly affects the quality of the fit given the width of the main sequence. The ages we have derived using the three different sets of isochrones are consistent with each other, as are the reddenings. The error on the distance modulus is difficult to determine given the width of the MS and $V$ mag range of the clump, but we have chosen 0.2 mag because that is the amount required to shift the isochrone vertically by the margin seen in the example fits to the MS in Fig. \ref{iso_mag}. This results in an error on the solar distance of 0.2 kpc, and an error on the Galactocentric distance of 0.15 kpc. None of the three models fit the member photometry perfectly, but the PARSEC models provide the best overall fit, so we adopt a cluster age of 0.8 $\pm$ 0.1 Gyr, an apparent distance modulus of 14.15 $\pm$ 0.20, and a solar distance of 2.10 $\pm$ 0.20 kpc.
\\
\indent These cluster parameters are consistent with those in the literature except for the reddening; the PARSEC E($B-V$) of 0.82$^{+0.11}_{-0.08}$ is consistent with the \citet{Carraro06} value but not the \citet{BB07} E($B-V$) of 0.58. As the cluster appears to have variable reddening, it is perhaps not surprising that the calculated reddening would vary depending on the fitting method. Even with member selection, the MS is still broad in ($J-H$) and the clump spreads across $\sim$0.7 mag in $J$. Fitting isochrones to $J$, $J-H$ and $J$, $J-K$ space gives little additional leverage on cluster parameters, partly because of the lack of narrowing in the clump and MS and partly because models predict less change in these colors with cluster age, so we do not utilize this parameter space. It is useful to note that since none of the studies of Tr 23 use $B$ photometry, these E($B-V$) values are subject to differences in color conversions. When fitting isochrones to the 2MASS photometry of members we observe an E($J-H$) of 0.21 for Tr 23 while \citet{BB07} find an E($J-H$) of 0.18; this is within the errors we would expect due to the size of the clump in ($J-H$) of $\sim$0.1 mag.
\\
\indent It would seem that the field of Tr 23 either has strong DR or the width of the MS is caused by effects unrelated to extinction. The width of the lower member MS in both ($V-I$) and ($J-H$) might be explained by binaries, as a variation in $V$ of 0.75 mag caused by pairs of equal-mass stars would encompass most member stars with $V$ below 16. However, there is a clump of stars with 15.5 $<$ $V$ $<$ 16 that are too far from the MS to be fully explained by binary stars. If we assume that the separation between the two apparent main sequences shown in Fig. \ref{diff_red} is entirely due to binaries, the binary fraction of the cluster would be at least 30\%. This would appear to require the maximum separation of 0.7 mag (i.e., equal mass stars) for most of the ten stars we have marked as the reddened MS, so the actual binary fraction of the cluster might be significantly higher. Observational estimates of the fraction of short-period ($<$ 10$^4$ days) binaries in OCs typically fall at $\sim$20\% \citep{Duchene99, Mermilliod08, Milliman14}; total binary fractions may be substantially higher. A mixture of DR and binary widening of the MS is perhaps the most likely scenario, and could explain well the full extent of the MS; however, the two effects are difficult to disentangle without more extensive cluster membership data.

\section{Abundance analysis} \label{abuns}

\begin{figure*}
\centering
\includegraphics{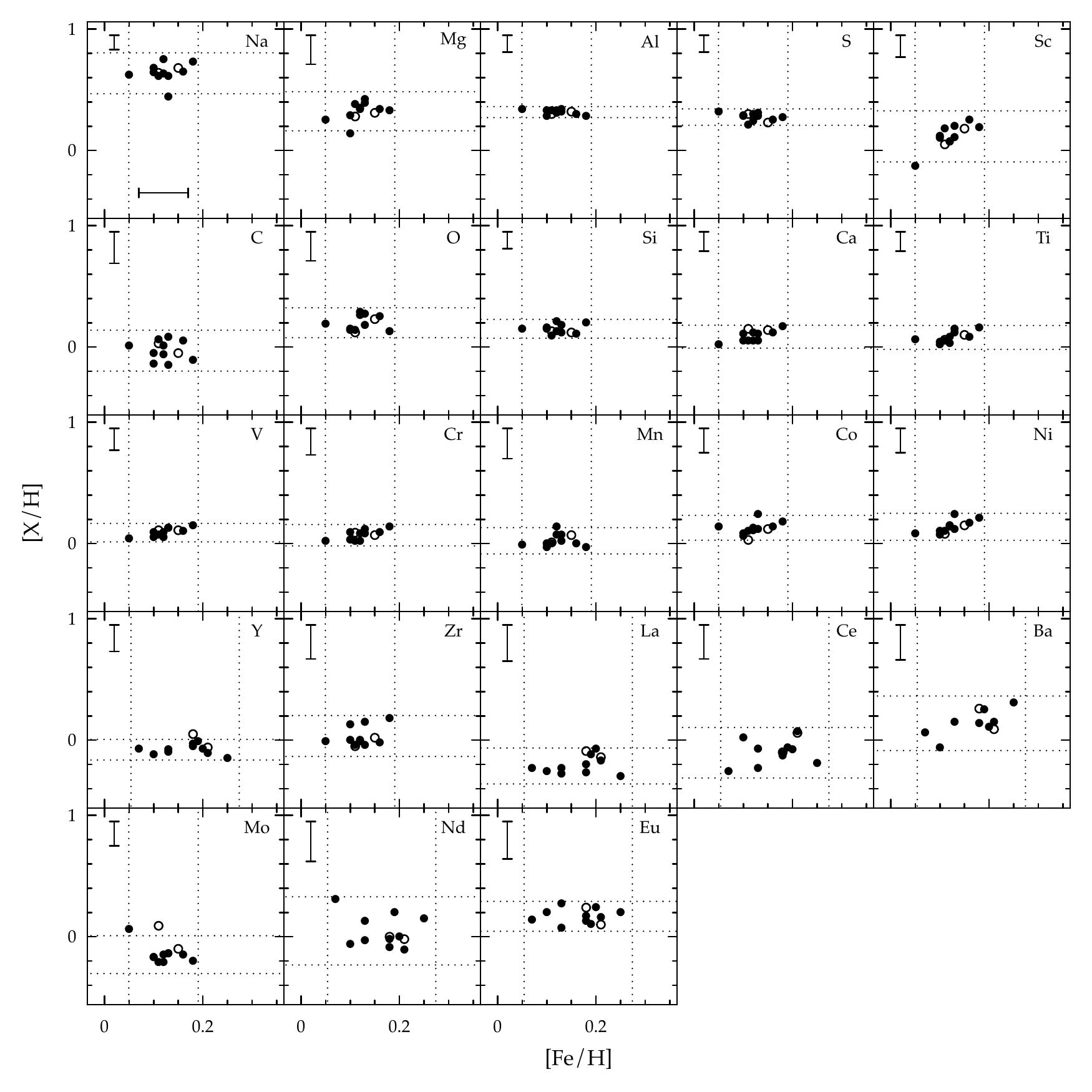}
\caption{Tr 23 stellar [X/H] abundances with representative stellar error bars. Filled circles are radial velocity members, and open circles are possible binary members. Dashed lines indicate 2$\sigma$ from the cluster average as calculated based on RV member abundances.}
\label{abun_fig}
\end{figure*}

\begin{figure*}
\centering
\includegraphics[scale=1.0]{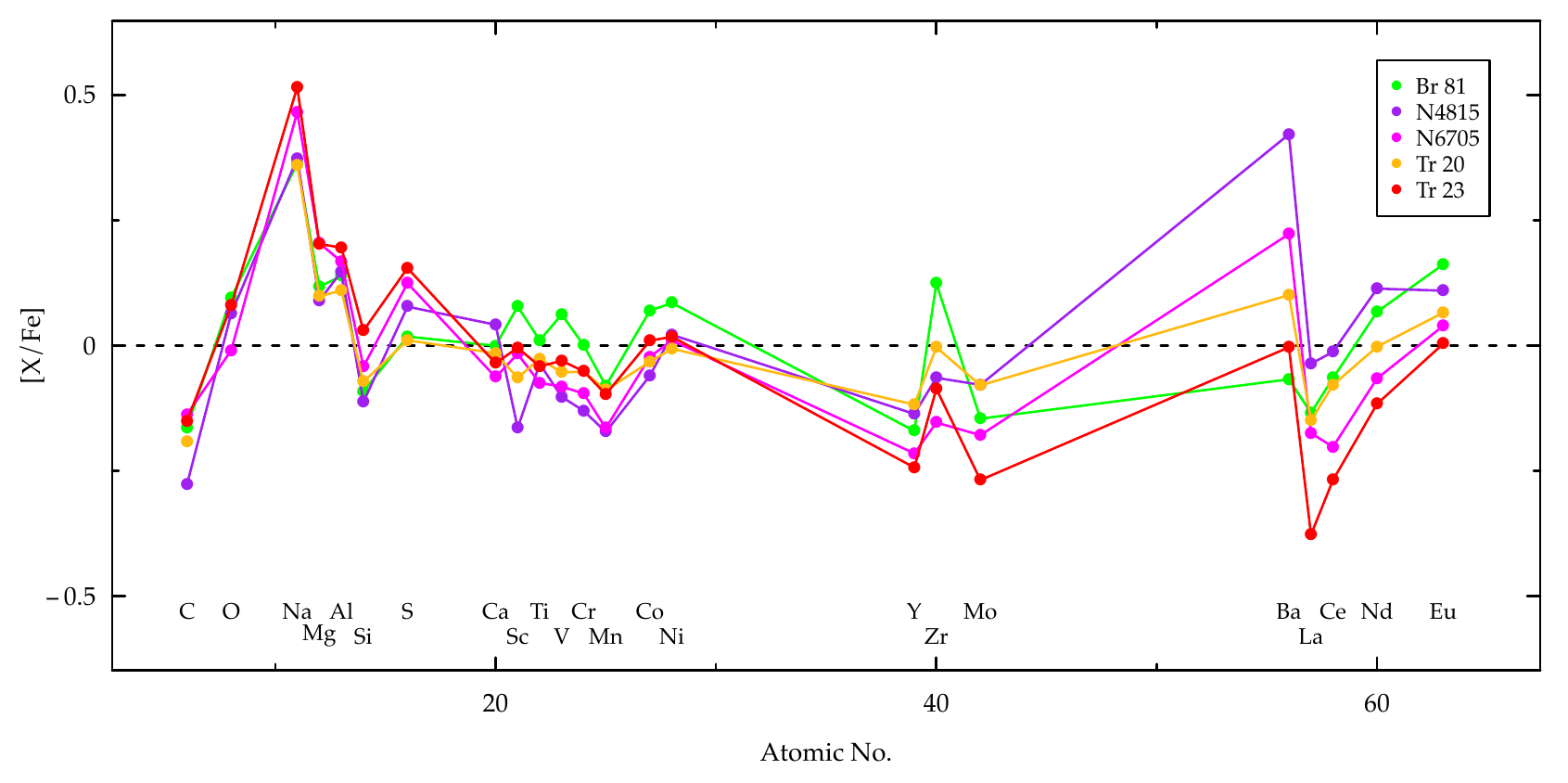}
\caption{Tr 23 and previously analyzed GES OC abundances vs. atomic number. All abundance ratios are taken from iDR4 analyses and normalized to \citet{Grevesse07} solar abundances.}
\label{atm_no}
\end{figure*}

Tables \ref{uves_abun1}, \ref{uves_abun2}, and \ref{uves_abun3} give stellar abundances of UVES cluster radial velocity members for light and $\alpha$ elements, Fe peak elements, and neutron-capture elements, respectively. The errors for each stellar measurement are based on the line-by-line abundance variations and differences in stellar abundances between abundance analysis nodes. Some abundance measurements (\ion{C}{I}) are based on the measurements of a single node but most involve contributions from multiple nodes. The stellar error is then calculated by taking the median absolute deviation (MAD) of different node measurements of line abundances. For a set of abundances for one line from different nodes, the absolute values of the differences between the individual values and the group median are combined and the medians of all of these values is taken as the stellar error. Table \ref{avg_abun} gives cluster averages, standard deviations, errors on each cluster mean (representative of internal errors), and the median stellar error on the abundance (more representative of external errors) for each species measured relative to iron. Reference solar abundances from \citet{Grevesse07} are also given.
\\
\indent The Tr 23 averages and dispersions are based on the ten UVES radial velocity members as indicated in Table \ref{uves_info}; we note that cluster standard deviations and errors on the mean are based on the stellar measurements of the element only, and do not include errors on Fe or errors due to atmospheric parameter errors. In the following discussion, we reference the standard deviations as errors on the given cluster average abundances since these more closely reflect the stellar abundance errors (due to different measurement techniques and line-by-line analysis as described in Section \ref{targets}).
\\
\indent Star 128 is about 300K cooler than the other nine clump stars that are radial velocity members. Its abundances differ from the other clump stars for several elements; it is 2$\sigma$ from the cluster average for C, Na, Co, and Ni. However, given the number of stars considered, we might expect one cluster member to be outside of the 2$\sigma$ boundary for at least some elements. Figure \ref{abun_fig} shows [X/H] ratios for clump stars vs. [Fe/H]. Members are marked as filled circles; stars 74 and 86 have velocities 2 to 3 $\sigma$ from the cluster mean but may be binary cluster members, so we mark them on the element distribution plots as open circles. These two stars do not distinguish themselves from the radial velocity member abundances in any clear way; star 86 is 2$\sigma$ from the cluster average \ion{Y}{II}, but again the deviation for a single element is not significant in itself. 
\\
\indent We do not see any evidence of a spread in cluster abundances for the elements measured; in Table \ref{avg_abun} the standard deviations of [X/Fe] ratios are smaller than the median stellar error for all elements except sodium. The standard deviation on the cluster [Na/Fe] abundance is slightly higher than the median stellar Na error, but this difference is probably not significant, as any real enhancements in stellar Na within the cluster should occur along the RGB. Assuming that all stars targeted with UVES are indeed clump stars, we do not expect to see large variations in [Na/Fe] as the stars are all in the same evolutionary state.
\\
\indent Figure \ref{atm_no} shows cluster averages for Tr 23 and four inner-disk OCs analyzed in previous GES papers vs. atomic number; agreement is generally good except for a few elements discussed in the following sections. We note that this figure is based on iDR4 abundances, with averages re-calculated from member stars selected in \citet{Friel14} (NGC 4815), \citet{CantatGaudin14} (NGC6705), \citet{Magrini15} (Be 81), and \citet{Donati14} (Tr 20).

\subsection{Light elements} \label{light_abs}

We find cluster averages for carbon and oxygen of --0.15 $\pm$ 0.09 dex and +0.08 $\pm$ 0.07 dex respectively. \citet{Taut10} find roughly solar [O/Fe] ratios, and consistent [C/Fe] ratios of $\sim$ -0.2 dex, for Galactic clump stars. The [Fe/H] and [C/Fe] abundances determined by GES for intermediate-age OC Trumpler 20 \citep{Taut15} at +0.10 and --0.20 dex are similar to those we find for Tr 23; abundances for NGC 4815 and NGC 6705, two clusters closer in age to Tr 23, have solar [Fe/H] abundances but also show subsolar [C/Fe] ratios of --0.16 and --0.08 dex, respectively. All three of these clusters fall at similar Galactocentric radii as Tr 23. As shown in \citet{Taut15}, models from \citet{Magrini09} and \citet{Romano10} predict solar or slightly subsolar [O/Fe] abundances for an R$_{\mathrm{GC}}$ = 6 kpc, although GES measurements for NGC 4815 and NGC 6705 [O/Fe] are +0.13 dex (Tr 20 stars could not be measured due to telluric contamination of the O feature). Unlike C, stellar evolution models predict that O abundances of evolved stars are primordial. The Tr 23 dispersion in [C/Fe] is slightly larger than [O/Fe], although as with Na we do not expect to observe any significant dispersion in the carbon abundance from clump stars alone.
\\
\indent Tr 23 is significantly enhanced in sodium, with an average measured [Na/Fe] = +0.52 $\pm$ 0.09 dex. We calculate NLTE corrections following \citet{Smiljanic16} and find a typical correction $<$Na$_{\mathrm{NLTE}}$ -- Na$_{\mathrm{LTE}}>$ = --0.10 dex for Tr 23 clump stars, and a cluster average [Na/Fe]$_{\mathrm{NLTE}}$ of +0.42 $\pm$ 0.08 dex. \citet{Smiljanic16} also study Na abundances for field dwarfs and giants and find that giant abundances are systematically higher than dwarf abundances for a range of metallicities. Some mixing of products of the NeNa cycle is expected during the first dredge-up stage \citep{EEC95}, so Na increases in giants may be a real result of stellar evolution, or partly real and partly due to systematic effects. Models of [Na/Fe] for different stellar TO masses from \citet{Lagarde12} predict an NLTE [Na/Fe] of +0.3 for solar [Fe/H] stars at M$_{\mathrm{TO}}$ = 3M$_{\odot}$, including the effects of stellar rotation. [Na/Fe] abundances for other GES OCs seem to match the model predictions, including the youngest of the selected GES clusters, NGC 6705, at 0.3 Gyr and [Na/Fe]$_{\mathrm{NLTE}}$ = 0.38 $\pm$ 0.11. The Tr 23 [Na/Fe] abundance is higher than the \citet{Lagarde12} rotational mixing models predict for its turnoff mass (M$_{\mathrm{TO}}$ = 2.1M$_{\odot}$; see Table \ref{iso_params}), but it falls within the giant field star and OC distributions in \citet{Smiljanic16} and follows the same general trend of increasing [Na/Fe] abundance with increasing TO mass.
\\
\indent We find an [Al/Fe] abundance of +0.20 $\pm$ 0.05 dex; compared to OC aluminum abundances in \citet{Smiljanic16}, which are based on a solar $\epsilon$(Al) = 6.44, the Tr 23 average is typical for solar-metallicity field giants, though it is on the upper edge of the dwarf star distributions in Smiljanic et al. (2016) and \citet{Bensby14}. Once the typical NLTE correction for giants from \citet{Smiljanic16} of -0.05 dex is applied, the cluster average on that scale becomes +0.08 dex, supporting their finding that OC giants are not enhanced in [Al/Fe] relative to model predictions of \citet{Lagarde12}.

\subsection{$\alpha$-elements} \label{alpha}

Tr 23 has roughly solar [$\alpha$/Fe] ratios for all elements except magnesium. We find an enhanced [Mg/Fe] = +0.20 $\pm$ 0.07, inconsistent with the cluster [Si/Fe], [Ca/Fe] and [Ti/Fe] of 0.03 $\pm$ 0.05, --0.04 $\pm$ 0.03, and --0.04 $\pm$ 0.04, respectively. Previous GES abundance studies have also found enhanced [Mg/Fe] in OCs \citep[see][]{Magrini14} compared to other $\alpha$-elements by about 0.2 dex. We note that three Mg lines at $\sim$6300$\mathrm{\AA}$ which are commonly measured via equivalent widths may be affected by a Ca autoionization feature that increases the measured abundances by about 0.2 dex in solar metallicity red giants \citep{Overbeek15}. However, \citet{Magrini15} note that O and Mg are almost entirely produced in type II supernovae while Si, Ca, and Ti can have considerable contributions (50 to 70\%) from SN Ia; Tr 23 abundances for these two groups are consistent with each other within errors.
\\
\indent $\alpha$-element measurements for the three previously mentioned inner disk GES OCs were presented in \citet{Magrini14}; they find solar or slightly subsolar [Si/Fe], [Ca/Fe], and [Ti/Fe] abundances consistent with solar metallicity field stars of the inner disk and bulge. Cepheid data from \citet{And02} also indicate that [$\alpha$/Fe] ratios in the inner disk are roughly solar and show little relationship with R$_{\mathrm{GC}}$.

\subsection{Fe peak elements}

Sc, V, Cr, Mn, Co, and Ni ratios to Fe are all roughly solar. The inner disk clusters also show little variation in these abundances.

\subsection{Neutron-capture elements}

\begin{figure*}
\centering
\includegraphics[scale=0.8]{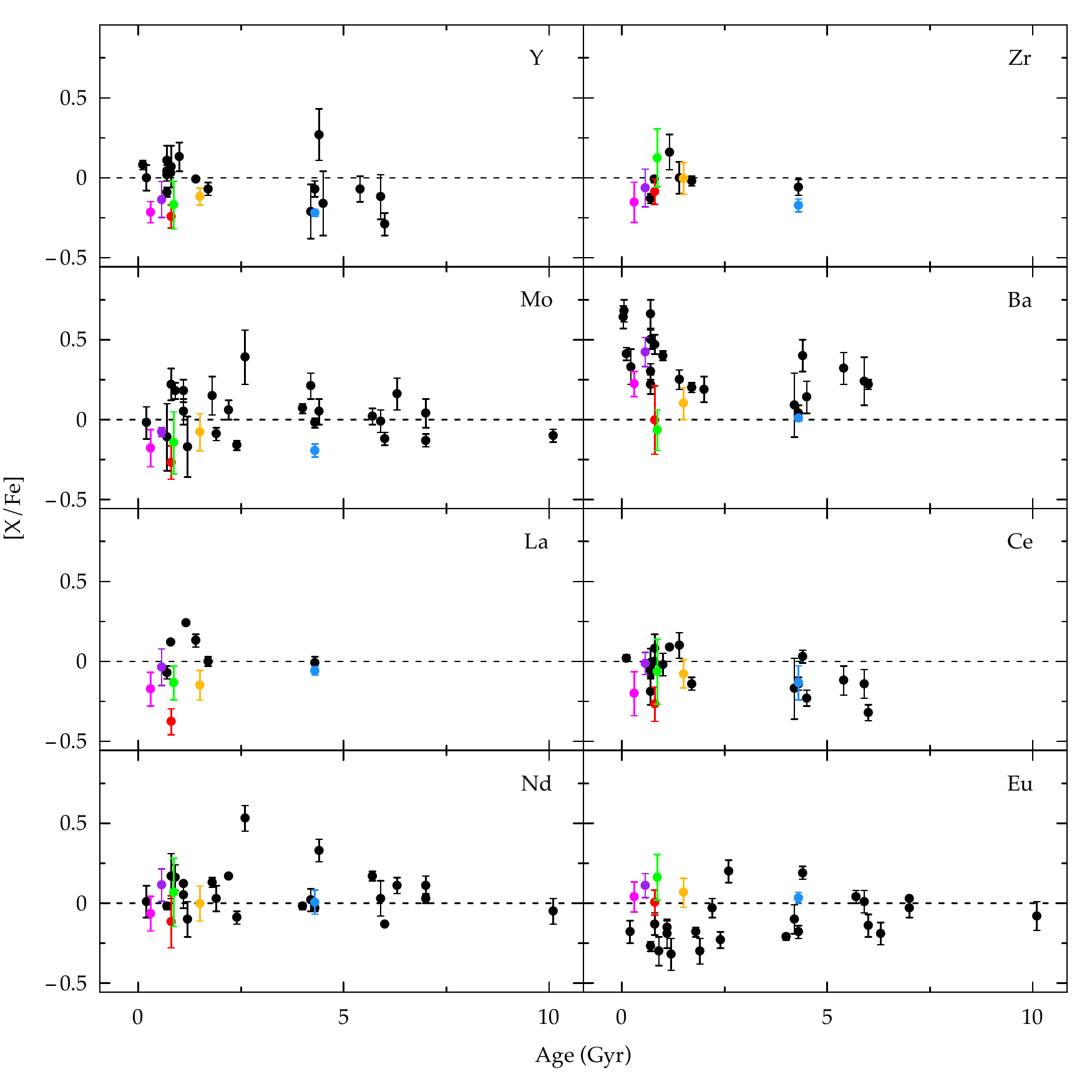}
\caption{Tr 23 [X/Fe] abundances vs. age in red and M67 abundances in blue compared to literature OC abundances from \citet{DOrazi09, Maiorca11, Overbeek16} in black and other GES OCs the same colors as Figure \ref{atm_no}.}
\label{OC_comp_age}
\end{figure*}

\begin{figure*}
\centering
\includegraphics[scale=0.8]{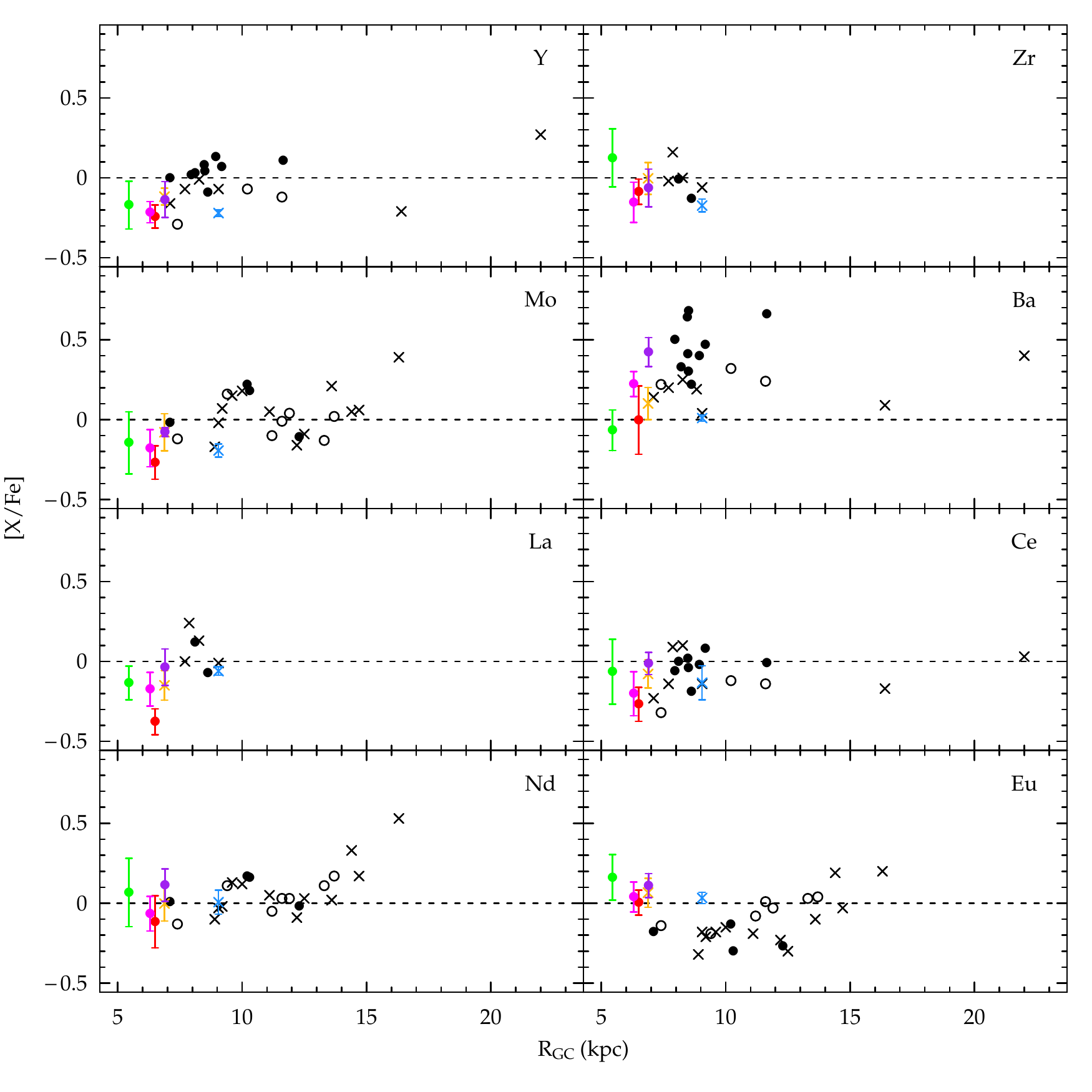}
\caption{Tr 23 [X/Fe] abundances vs. Galactocentric radius; colors are the same as in Figure \ref{OC_comp_age}. Filled circles are clusters $<$ 1 Gyr old, crosses are clusters 2 to 5 Gyr old, and open circles are clusters $>$ 5 Gyr old.}
\label{OC_comp_rgc}
\end{figure*}

Y, Zr, Mo, Ba, La, and Ce are majority slow neutron-capture (s-process) elements. There are multiple nucleosynthesis pathways for s-process elements \citep[see, e.g.,][]{Travaglio04}, but AGB stars are thought to be the most significant contributors for young clusters. There is a continuing debate about the size of the contribution from AGB stars; supersolar abundances measured in young OCs for some elements, primarily Ba, suggest that the role of low-mass AGB stars is larger than once thought \citep{DOrazi09, Maiorca11, Mishenina15}. However, other s-process elements do not always show the same enhancement. \citet{JF13} measured [Zr/Fe] and [La/Fe] abundances in a sample of 19 OCs and found that the scatter in cluster abundance, particularly for ages $<$ 3 Gyr, is larger than the magnitude of the trends in these elements with age.
\\
\indent OCs younger than 3 Gyr are generally found to have solar or supersolar s-process abundances \citep{Maiorca11}; Tr 23, however, is subsolar in all measured s-process elements except Ba. We may further subdivide s-process elements into light (Y, Zr, Mo) and heavy (Ba, La, Ce); each of these groups should have very similar abundances. The neutron flux inside AGB stars, as well as the metallicity (i.e., the number of Fe 'seeds' for neutrons to build on), may affect the relative abundances of the light vs. heavy elements, but the abundances of nuclei of similar atomic weight should strongly correlate. We plot GES cluster averages for neutron-capture elements with age along with literature data from \citet{Maiorca11} (Y, Zr, La, and Ce), \citet{DOrazi09} (Ba), and \citet{Overbeek16} (Mo, Nd, and Eu) in Figure \ref{OC_comp_age}. Colors for GES OCs in Figure \ref{OC_comp_age} are the same as Figure \ref{atm_no} except that M67 is marked in blue (the M67 average is based on giant stars only). The error bars in Figure \ref{OC_comp_age} represent the standard deviations of stellar abundances; the errors on mean cluster abundances are sometimes smaller than the points, particularly for Tr 20 (42 member stars) and N6705 (27 member stars). All literature abundances have been placed on the solar abundance scale of \citet{Grevesse07}. We see in Figure \ref{OC_comp_age} that the GES clusters, and particularly Tr 23, have lower s-process abundances than literature clusters of the same ages. If we look specifically at M67, which is represented in Figure \ref{OC_comp_age} in the literature sample as well as the GES sample (at 4.3 Gyr), we see systematic offsets between the GES and literature abundances, particularly for the light s-process elements. The GES M67 abundances are lower than found by \citet{Maiorca11} and \citet{Overbeek16} by 0.1 to 0.2 dex for the light s-process elements, and if we consider these as indicative of systematic offsets between the GES and literature abundance scales, the GES clusters are not significantly lower in light s-process elements than literature clusters of the same ages. However, the M67 heavy s-process element abundances of the two samples agree well, and the younger GES OCs have lower average abundances than literature clusters in their age group. Tr 23 in particular falls below the range of literature abundances for Ba, La, and Ce. We also see a suggestion of large [Ba/Fe] increases ($\sim$0.4 dex) for the youngest two GES clusters (N4815 and N6705) relative to the older clusters, which is not reflected in La or Ce abundances, or the light s-process elements. 
\\
\indent In Figure \ref{OC_comp_rgc} we plot the same GES and literature data against Galactocentric radius using the same colors and abundance error bars as in Figure \ref{OC_comp_age}, but with point types marking different age groups; filled circles are clusters $\le$ 1 Gyr old, crosses are clusters between 1 and 5 Gyr, and open circles are clusters older than 5 Gyr. Looking at all age groups together, we see no clear sign of a relationship between [s-/Fe] abundances and Galactocentric radius. The literature [Zr/Fe] and [La/Fe] samples are not large enough for definite interpretation, but when the systematic differences between M67 values are taken into account, the inner disk (GES) clusters are at the same [Y/Fe], [Mo/Fe], and [Ce/Fe] abundances as literature clusters in the solar neighborhood. [Ba/Fe] shows suggestions of a decrease in the inner disk based on the GES cluster abundances, but in the larger \citet{DOrazi09} sample there is considerable scatter around the solar neighborhood, so more data points would be needed in the inner disk to confirm the decrease. When looking at the individual age groups, other than Ba abundances in young OCs there does not seem to be much of a gradient in the light or heavy s-process elements for any of the three age bins.
\\
\indent Nd is a mixed r- and s-process element, and Eu is a majority r-process element, with 42 and 98\% solar abundance formed via the r-process, respectively \citep{Sneden08}. r-process elements are formed in high neutron flux environments, most likely type II supernovae or neutron star mergers. Because these involve different mass ranges of stars than the s-process we necessarily expect relationships of r-process element abundances with cluster age to be different than for the s-process. For Tr 23, we find a solar [Eu/Fe] ratio of 0.00 $\pm$ 0.08 dex and slightly subsolar [Nd/Fe] of -0.12 $\pm$ 0.16. These are similar to the other GES OC abundances in Figure \ref{atm_no}, although Tr23 still has the lowest abundances for both elements. In Figure \ref{OC_comp_age} we see that the Tr 23 [Nd/Fe] abundance is within the literature OC distribution although it is on the low end; the other GES clusters match the literature Nd distribution well. The [Eu/Fe] abundances for Tr 23 and the other GES OCs are higher than literature abundances, but if we correct for the apparent offset in measured M67 Eu abundances, the GES OCs are only slightly enhanced in Eu relative to the literature abundances for young clusters. 
\\
\indent In Figure \ref{OC_comp_rgc}, [Nd/Fe] does seem to increase with Galactocentric radius, largely due to clusters more than 14 kpc from the Galactic center and especially Be 31 (at 16 kpc) which is based on a single star displaying unusual abundance patterns \citep{Overbeek16, Yong05}. For [Eu/Fe], the intermediate age clusters do show some increase with R$_{\mathrm{GC}}$; the youngest clusters do not appear to have a gradient, but they also do not probe the outer regions of the disk (beyond 13 kpc) where we see significant increases for older clusters.
\\
\indent It would seem from these data that the inner disk clusters are modestly enhanced in Eu and significantly deficient in heavy s-process element abundances relative to other OCs of similar ages. The scatter in the inner disk clusters, however, is similar to scatter in young clusters in the literature for both r- and s-process elements. The scatter observed in neutron-capture abundances among the GES OCs of similar ages and similar Galactocentric radii (Br 81, N4815, N6705, and Tr 23 are all younger than 1 Gyr and within 7 kpc of the Galactic center) is significant given the small errors on the means; this scatter may be due to localized abundance inhomogeneities in the interstellar medium (ISM). \citet{Maiorca12} model s-process yields for AGB stars of different masses and different mass C$^{13}$ pockets (the primary source of neutrons in low-mass AGB stars) and show that moderately metal-poor AGB stars of 2 M$_{\odot}$ overproduce light s-process elements relative to heavy s-process elements, whereas a star of 1.5 M$_{\odot}$ with a larger C$^{13}$ pocket would produce higher yields of heavy s-process elements than light s-process elements. They further note that the factors determining the size of the C$^{13}$ pocket in low-mass stars are not well understood or constrained by observations; it may be that the gas from which Tr 23 formed was preferentially enriched by higher-mass AGB stars which left it deficient in heavy s-process elements, although we would typically expect the inner disk to have old enough stellar populations so that very low-mass AGB stars could have already enriched the ISM. It is also possible that different low-mass AGB stars might have different C$^{13}$ pocket masses (the physical factors driving the pocket mass are unclear so there could be some variation) and produce the light and heavy s-process elements with different efficiencies. 
\\
\indent Radial migration may also play a role in moderating any existing gradients and increasing scatter at different Galactocentric radii and ages. Most clusters, however, are not expected to move more than a few kpc and likely occur primarily in the outer disk \citep{Bird12}, so this cannot be the sole explanation for the behavior of heavy s-process elements in Tr 23.

\section{Summary and conclusions} \label{summary}

The Gaia-ESO Survey has provided the first spectroscopy of the inner disk open cluster Trumpler 23, which has only two existing photometry studies \citep{Carraro06, BB07} -- it is only 20 pc above the plane of the Galaxy. We use Gaia-ESO radial velocity measurements to find a cluster systemic radial velocity of -61.3 $\pm$ 1.9 \kms~(s.d.), and isolate 70 stars out of 167 observed as radial velocity members. We are also able to use GES atmospheric parameters to remove a field star present in the selected cluster radial velocity distribution. Based on our radial velocity membership criteria, we derive an [Fe/H] of +0.14 $\pm$ 0.03 consistent with other OCs in this area of the disk.
\\
\indent Using only stars we have selected as radial velocity members, we re-determine cluster parameters by fitting PARSEC, BaSTI, and Dartmouth isochrones to $V$, $V-I$ photometry from \citet{Carraro06}. With these three sets of model isochrones, we derive ages from 0.60 to 0.80 Gyr, E($V-I$) from 1.02 to 1.10, and distance moduli from 14.15 to 14.50. We adopt the PARSEC isochrone parameters of 0.80 $\pm$ 0.10 Gyr, E($V-I$) = 1.02$^{+0.14}_{-0.09}$ (E($B-V$) = 0.82$^{+0.11}_{-0.08}$), and (m$-$M)$_{\mathrm{V}}$ = 14.15 $\pm$ 0.20 (d = 2.10 $\pm$ 0.20, R$_{\mathrm{GC}}$ =  6.30). Our cluster parameters agree within the errors with determinations from both previous photometry studies except for the \citet{BB07} E($B-V$) which is substantially lower than ours at 0.58 $\pm$ 0.03. However, the \citet{BB07} reddening is based on isochrone fitting to 2MASS $J,H,K$ photometry, and the color relations used may be the source of the discrepancy. We find an apparent spread in the lower MS which is likely due to broadening from both DR and binaries.
\\
\indent We also present ten member star abundances for 23 elements plus iron, based on high-resolution UVES spectra. We find that the cluster has an approximately solar [$\alpha$/Fe] ratio, though with enhanced [Mg/Fe]. It also has an unusually high [Na/Fe] of 0.42 $\pm$ 0.08 which is not seen in other intermediate-age inner disk open clusters analyzed by GES; it does however generally agree with previous findings of increasing [Na/Fe] abundances with increasing cluster turn-off mass. [C/Fe] and [O/Fe] abundances are typical of open clusters; because our high-resolution data was taken for clump stars only we cannot examine changes in C or Na with evolutionary state, and we find no evidence of internal spread in the cluster for any elements. Tr 23 has a solar r-process ratio (as measured by Eu) but appears to be deficient in s-process elements, particularly La, Ce, and Ba; this may be due in part to radial migration, localized enrichment of the ISM, or varying efficiencies of neutron sources in AGB stars.

\begin{acknowledgements}

Based on data products from observations made with ESO Telescopes at the La Silla Paranal Observatory under programme ID 188.B-3002. These data products have been processed by the Cambridge Astronomy Survey Unit (CASU) at the Institute of Astronomy, University of Cambridge, and by the FLAMES/UVES reduction team at INAF/Osservatorio Astrofisico di Arcetri. These data have been obtained from the Gaia-ESO Survey Data Archive, prepared and hosted by the Wide Field Astronomy Unit, Institute for Astronomy, University of Edinburgh, which is funded by the UK Science and Technology Facilities Council.

This work was partly supported by the European Union FP7 programme through ERC grant number 320360 and by the Leverhulme Trust through grant RPG-2012-541. We acknowledge the support from INAF and Ministero dell' Istruzione, dell' Universit\`a' e della Ricerca (MIUR) in the form of the grant "Premiale VLT 2012". The results presented here benefit from discussions held during the Gaia-ESO workshops and conferences supported by the ESF (European Science Foundation) through the GREAT Research Network Programme.

The research leading to these results has received funding from the European Community's Seventh Framework Programme (FP7-SPACE-2013-1) under grant agreement no. 606740.

D.G., S.V., B.T., and C.M. gratefully acknowledge support from the Chilean BASAL Centro de Excelencia en Astrof\'isica y Tecnolog\'ias Afines (CATA) grant PFB-06/2007.

\end{acknowledgements}

\bibliographystyle{aa}
\bibliography{Tr23_refs}

\begin{table*}
\caption{Parameters for Trumpler 23 Radial Velocity Members with GIRAFFE Data \label{gir_info}}
\setlength{\tabcolsep}{0.05in}
\centering
\scriptsize
\begin{tabular}{l c c c c c l r r r r}  
\hline\hline
ID$^a$ & GES ID & $V^a$ & $V-I^a$ & RA & Dec. & T$_{\mathrm{eff}}$ & log(g) & $\xi$ & [Fe/H] & V$_\mathrm{r}$ \\ & & (mag) & (mag) & (deg.) & (deg.) & (K) & (cm s$^{-2}$) & (km s$^{-1}$) & (dex) & (km s$^{-1}$) \\
\hline
18 & 16005705-5333242 & 13.950 & 2.179 & 240.23771 & -53.55672 & 5501 $\pm$ 150 & 4.42 $\pm$ 0.02 & 1.12 $\pm$ 0.02 & 0.65 $\pm$ 0.29 & -62.86 $\pm$ 0.10 \\
19 & 16004034-5333240 & 13.938 & 2.271 & 240.16808 & -53.55667 & $\ldots$ & $\ldots$ & $\ldots$ & $\ldots$ & -61.17 $\pm$ 0.10 \\
42 & 16005086-5332030 & 14.229 & 1.673 & 240.21192 & -53.53417 & 6063 $\pm$ 145 & 3.17 $\pm$ 0.03 & 1.58 $\pm$ 0.05 & $\ldots$ & -62.63 $\pm$ 0.10 \\
49 & 16003972-5331217 & 14.277 & 1.607 & 240.16550 & -53.52269 & $\ldots$ & $\ldots$ & $\ldots$ & $\ldots$ & -59.64 $\pm$ 0.13 \\
112 & 16004929-5331023 & 14.717 & 1.352 & 240.20538 & -53.51731 & $\ldots$ & $\ldots$ & $\ldots$ & $\ldots$ & -62.53 $\pm$ 0.16 \\
141 & 16005986-5329431 & 14.889 & 1.207 & 240.24942 & -53.49531 & $\ldots$ & $\ldots$ & $\ldots$ & $\ldots$ & -62.43 $\pm$ 0.18 \\
149 & 16010227-5333466 & 15.013 & 1.329 & 240.25946 & -53.56294 & $\ldots$ & $\ldots$ & $\ldots$ & $\ldots$ & -64.65 $\pm$ 1.27 \\
165 & 16004809-5332301 & 15.123 & 1.251 & 240.20038 & -53.54169 & $\ldots$ & $\ldots$ & $\ldots$ & $\ldots$ & -64.21 $\pm$ 0.11 \\
172 & 16005462-5330451 & 14.998 & 1.234 & 240.22758 & -53.51253 & $\ldots$ & $\ldots$ & $\ldots$ & $\ldots$ & -62.72 $\pm$ 0.21 \\
179 & 16004639-5333149 & 15.263 & 1.259 & 240.19329 & -53.55414 & $\ldots$ & $\ldots$ & $\ldots$ & $\ldots$ & -62.81 $\pm$ 0.51 \\
197 & 16005412-5336140 & 15.735 & 2.141 & 240.22550 & -53.60389 & $\ldots$ & $\ldots$ & $\ldots$ & $\ldots$ & -64.29 $\pm$ 0.11 \\
199 & 16010556-5333133 & 15.245 & 1.208 & 240.27317 & -53.55369 & $\ldots$ & $\ldots$ & $\ldots$ & $\ldots$ & -60.00 $\pm$ 0.50 \\
246 & 16004693-5334523 & 15.460 & 1.210 & 240.19554 & -53.58119 & $\ldots$ & $\ldots$ & $\ldots$ & $\ldots$ & -64.70 $\pm$ 0.32 \\
254 & 16010935-5332003 & 15.410 & 1.196 & 240.28896 & -53.53342 & $\ldots$ & $\ldots$ & $\ldots$ & $\ldots$ & -61.79 $\pm$ 0.68 \\
257 & 16004424-5332421 & 15.552 & 1.360 & 240.18433 & -53.54503 & 7530 $\pm$ 44 & $\ldots$ & $\ldots$ & $\ldots$ & -60.12 $\pm$ 0.25 \\
260 & 16004496-5333060 & 15.567 & 1.378 & 240.18733 & -53.55167 & $\ldots$ & $\ldots$ & $\ldots$ & $\ldots$ & -63.87 $\pm$ 1.72 \\
286 & 16005715-5332459 & 15.654 & 1.361 & 240.23812 & -53.54608 & $\ldots$ & $\ldots$ & $\ldots$ & $\ldots$ & -57.42 $\pm$ 2.13 \\
297 & 16005224-5331009 & 15.647 & 1.328 & 240.21767 & -53.51692 & 7755 $\pm$ 52 & $\ldots$ & $\ldots$ & $\ldots$ & -63.40 $\pm$ 0.45 \\
301 & 16004889-5331273 & 15.717 & 1.335 & 240.20371 & -53.52425 & 7621 $\pm$ 54 & $\ldots$ & $\ldots$ & $\ldots$ & -60.65 $\pm$ 0.34 \\
312 & 16005370-5333549 & 15.679 & 1.210 & 240.22375 & -53.56525 & 7924 $\pm$ 63 & $\ldots$ & $\ldots$ & $\ldots$ & -62.62 $\pm$ 0.42 \\
319 & 16003077-5331541 & 15.860 & 1.412 & 240.12821 & -53.53169 & 6039 $\pm$ 83 & 4.05 $\pm$ 0.26 & $\ldots$ & 0.04 $\pm$ 0.20 & -58.00 $\pm$ 0.24 \\
320 & 16005689-5332277 & 15.860 & 1.364 & 240.23704 & -53.54103 & $\ldots$ & $\ldots$ & $\ldots$ & $\ldots$ & -61.01 $\pm$ 1.22 \\
379 & 16010382-5331502 & 15.855 & 1.258 & 240.26592 & -53.53061 & 7987 $\pm$ 60 & $\ldots$ & $\ldots$ & $\ldots$ & -59.56 $\pm$ 0.25 \\
386 & 16005638-5333098 & 15.926 & 1.238 & 240.23492 & -53.55272 & 7852 $\pm$ 61 & $\ldots$ & $\ldots$ & $\ldots$ & -63.87 $\pm$ 0.49 \\
391 & 16004787-5334169 & 15.952 & 1.219 & 240.19946 & -53.57136 & $\ldots$ & $\ldots$ & $\ldots$ & $\ldots$ & -59.38 $\pm$ 2.02 \\
392 & 16005131-5333440 & 16.016 & 1.403 & 240.21379 & -53.56222 & $\ldots$ & $\ldots$ & $\ldots$ & $\ldots$ & -58.80 $\pm$ 0.48 \\
393 & 16003266-5329243 & 15.964 & 1.222 & 240.13608 & -53.49008 & $\ldots$ & $\ldots$ & $\ldots$ & $\ldots$ & -61.24 $\pm$ 1.94 \\
401 & 16004946-5329113 & 15.937 & 1.268 & 240.20608 & -53.48647 & $\ldots$ & $\ldots$ & $\ldots$ & $\ldots$ & -57.41 $\pm$ 7.53 \\
406 & 16004295-5330566 & 16.153 & 1.344 & 240.17896 & -53.51572 & $\ldots$ & $\ldots$ & $\ldots$ & $\ldots$ & -60.35 $\pm$ 2.04 \\
413 & 16010069-5329384 & 16.040 & 1.288 & 240.25287 & -53.49400 & $\ldots$ & $\ldots$ & $\ldots$ & $\ldots$ & -61.17 $\pm$ 4.45 \\
419 & 16005951-5335225 & 16.151 & 1.337 & 240.24796 & -53.58958 & $\ldots$ & $\ldots$ & $\ldots$ & $\ldots$ & -62.60 $\pm$ 3.36 \\
429 & 16004544-5333062 & 16.146 & 1.292 & 240.18933 & -53.55172 & $\ldots$ & $\ldots$ & $\ldots$ & $\ldots$ & -61.19 $\pm$ 3.68 \\
478 & 16010204-5332060 & 16.232 & 1.352 & 240.25850 & -53.53500 & $\ldots$ & $\ldots$ & $\ldots$ & $\ldots$ & -62.07 $\pm$ 2.28 \\
484 & 16005062-5331383 & 16.208 & 1.228 & 240.21092 & -53.52731 & $\ldots$ & $\ldots$ & $\ldots$ & $\ldots$ & -62.26 $\pm$ 0.39 \\
489 & 16010489-5329440 & 16.270 & 1.347 & 240.27038 & -53.49556 & $\ldots$ & $\ldots$ & $\ldots$ & $\ldots$ & -61.10 $\pm$ 0.53 \\
496 & 16010476-5328190 & 16.273 & 1.389 & 240.26983 & -53.47194 & $\ldots$ & $\ldots$ & $\ldots$ & $\ldots$ & -62.97 $\pm$ 1.91 \\
501 & 16010497-5333409 & 16.240 & 1.220 & 240.27071 & -53.56136 & 8177 $\pm$ 82 & $\ldots$ & $\ldots$ & $\ldots$ & -61.24 $\pm$ 0.32 \\
509 & 16004168-5332182 & 16.393 & 1.311 & 240.17367 & -53.53839 & $\ldots$ & $\ldots$ & $\ldots$ & $\ldots$ & -62.82 $\pm$ 0.53 \\
531 & 16005584-5328374 & 16.480 & 1.441 & 240.23267 & -53.47706 & 7712 $\pm$ 79 & $\ldots$ & $\ldots$ & $\ldots$ & -58.26 $\pm$ 0.76 \\
556 & 16010454-5328438 & 16.387 & 1.254 & 240.26892 & -53.47883 & $\ldots$ & $\ldots$ & $\ldots$ & $\ldots$ & -61.19 $\pm$ 0.86 \\
576 & 16005741-5335518 & 16.547 & 1.392 & 240.23921 & -53.59772 & 8021 $\pm$ 84 & $\ldots$ & $\ldots$ & $\ldots$ & -57.85 $\pm$ 0.50 \\
579 & 16010723-5333561 & 16.485 & 1.217 & 240.28012 & -53.56558 & 7582 $\pm$ 82 & $\ldots$ & $\ldots$ & $\ldots$ & -60.58 $\pm$ 0.46 \\
623 & 16010108-5331400 & 16.550 & 1.335 & 240.25450 & -53.52778 & 7604 $\pm$ 249 & 4.18 $\pm$ 0.15 & $\ldots$ & 0.02 $\pm$ 0.20 & -60.06 $\pm$ 0.46 \\
634 & 16005175-5332339 & 16.682 & 1.471 & 240.21562 & -53.54275 & 7306 $\pm$ 783 & 4.15 $\pm$ 0.18 & $\ldots$ & -0.07 $\pm$ 0.46 & -64.20 $\pm$ 0.57 \\
792 & 16005774-5334375 & 16.915 & 1.369 & 240.24058 & -53.57708 & 6812 $\pm$ 307 & 4.06 $\pm$ 0.19 & $\ldots$ & 0.19 $\pm$ 0.30 & -59.04 $\pm$ 0.58 \\
815 & 16004357-5328513 & 17.005 & 1.514 & 240.18154 & -53.48092 & 6722 $\pm$ 435 & 4.13 $\pm$ 0.02 & $\ldots$ & 0.32 $\pm$ 0.35 & -62.09 $\pm$ 0.53 \\
821 & 16004569-5332567 & 17.063 & 1.497 & 240.19037 & -53.54908 & 7058 $\pm$ 488 & 4.15 $\pm$ 0.21 & $\ldots$ & 0.11 $\pm$ 0.15 & -64.93 $\pm$ 0.66 \\
877 & 16004906-5330204 & 17.137 & 1.436 & 240.20442 & -53.50567 & 6809 $\pm$ 344 & 4.13 $\pm$ 0.19 & $\ldots$ & 0.25 $\pm$ 0.36 & -58.10 $\pm$ 0.97 \\
922 & 16005585-5335245 & 17.148 & 1.403 & 240.23271 & -53.59014 & 7062 $\pm$ 553 & 4.10 $\pm$ 0.19 & $\ldots$ & -0.07 $\pm$ 0.46 & -61.26 $\pm$ 0.70 \\
981 & 16004581-5331557 & 17.266 & 1.476 & 240.19088 & -53.53214 & 6766 $\pm$ 407 & 4.46 $\pm$ 0.36 & $\ldots$ & 0.20 $\pm$ 0.40 & -64.23 $\pm$ 0.98 \\
1004 & 16005827-5330323 & 17.325 & 1.449 & 240.24279 & -53.50897 & 7055 $\pm$ 666 & 4.05 $\pm$ 0.23 & $\ldots$ & 0.12 $\pm$ 0.14 & -60.44 $\pm$ 0.36 \\
1012 & 16010533-5333195 & 17.311 & 1.535 & 240.27221 & -53.55542 & 6378 $\pm$ 181 & 4.32 $\pm$ 0.27 & $\ldots$ & 0.17 $\pm$ 0.18 & -60.91 $\pm$ 0.32 \\
1016 & 16004350-5331482 & 17.327 & 1.429 & 240.18125 & -53.53006 & 6584 $\pm$ 440 & 4.05 $\pm$ 0.21 & $\ldots$ & 0.30 $\pm$ 0.42 & -60.13 $\pm$ 1.01 \\
1061 & 16004174-5333545 & 17.432 & 1.555 & 240.17392 & -53.56514 & 6406 $\pm$ 186 & 4.24 $\pm$ 0.32 & $\ldots$ & 0.06 $\pm$ 0.15 & -58.33 $\pm$ 0.29 \\
1355 & 16005469-5334324 & 17.828 & 1.647 & 240.22788 & -53.57567 & 5993 $\pm$ 156 & 4.18 $\pm$ 0.42 & $\ldots$ & -0.02 $\pm$ 0.21 & -57.32 $\pm$ 0.32 \\
1358 & 16005146-5332497 & 17.773 & 1.538 & 240.21442 & -53.54714 & 6517 $\pm$ 397 & 3.90 $\pm$ 0.16 & $\ldots$ & 0.27 $\pm$ 0.28 & -60.99 $\pm$ 1.09 \\
1436 & 16003891-5330304 & 17.913 & 1.629 & 240.16213 & -53.50844 & 6364 $\pm$ 181 & 4.64 $\pm$ 0.83 & $\ldots$ & 0.25 $\pm$ 0.20 & -60.30 $\pm$ 0.77 \\
1483 & 16005675-5335029 & 17.896 & 1.560 & 240.23646 & -53.58414 & 6527 $\pm$ 203 & 4.36 $\pm$ 0.43 & $\ldots$ & 0.25 $\pm$ 0.25 & -60.39 $\pm$ 0.45 \\
1491 & 16010323-5330273 & 17.938 & 1.556 & 240.26346 & -53.50758 & 6647 $\pm$ 492 & 4.05 $\pm$ 0.12 & $\ldots$ & 0.39 $\pm$ 0.43 & -61.07 $\pm$ 0.36 \\
1528 & 16005241-5331345 & 17.940 & 1.567 & 240.21838 & -53.52625 & 6602 $\pm$ 414 & 4.07 $\pm$ 0.19 & $\ldots$ & 0.27 $\pm$ 0.38 & -63.36 $\pm$ 0.61 \\
\hline
\end{tabular}
\tablefoot{a. ID numbers and photometry from \citet{Carraro06}.}
\end{table*}

\begin{table*}
\caption{Parameters for Trumpler 23 Radial Velocity Non-members with GIRAFFE Data \label{gir_nm_info}}
\setlength{\tabcolsep}{0.05in}
\centering
\scriptsize
\begin{tabular}{l c c c c c l r r r r}
\hline\hline
ID$^a$ & GES ID & $V^a$ & $V-I^a$ & RA & Dec. & T$_{\mathrm{eff}}$ & log(g) & $\xi$ & [Fe/H] & V$_\mathrm{r}$ \\  & & (mag) & (mag) & (deg.) & (deg.)& (K) & (cm s$^{-2}$) & (km s$^{-1}$) & (dex) & (km s$^{-1}$) \\
\hline
36 & 16005835-5329550 & 13.553 & 0.988 & 240.24312 & -53.49861 & $\ldots$ & $\ldots$ & $\ldots$ & $\ldots$ & -54.36 $\pm$ 0.11 \\
39 & 16004521-5332044 & 13.938 & 1.249 & 240.18838 & -53.53456 & $\ldots$ & $\ldots$ & $\ldots$ & $\ldots$ & -132.95 $\pm$ 0.10 \\
71 & 16004917-5333257 & 14.151 & 0.998 & 240.20487 & -53.55714 & $\ldots$ & $\ldots$ & $\ldots$ & $\ldots$ & 41.98 $\pm$ 0.61 \\
73 & 16004353-5330424 & 14.210 & 1.132 & 240.18138 & -53.51178 & $\ldots$ & $\ldots$ & $\ldots$ & $\ldots$ & -10.06 $\pm$ 0.10 \\
80 & 16005170-5333015 & 14.133 & 0.889 & 240.21542 & -53.55042 & $\ldots$ & $\ldots$ & $\ldots$ & $\ldots$ & -20.20 $\pm$ 0.11 \\
104 & 16003549-5331214 & 14.552 & 1.026 & 240.14787 & -53.52261 & $\ldots$ & $\ldots$ & $\ldots$ & $\ldots$ & -38.57 $\pm$ 0.50 \\
129 & 16003881-5331400 & 14.806 & 1.253 & 240.16171 & -53.52778 & $\ldots$ & $\ldots$ & $\ldots$ & $\ldots$ & -38.33 $\pm$ 0.11 \\
135 & 16004252-5330337 & 14.568 & 0.921 & 240.17717 & -53.50936 & $\ldots$ & $\ldots$ & $\ldots$ & $\ldots$ & -32.38 $\pm$ 0.54 \\
145 & 16010286-5328087 & 15.213 & 1.674 & 240.26192 & -53.46908 & $\ldots$ & $\ldots$ & $\ldots$ & $\ldots$ & -32.90 $\pm$ 0.17 \\
146 & 16003985-5333514 & 14.988 & 1.259 & 240.16604 & -53.56428 & $\ldots$ & $\ldots$ & $\ldots$ & $\ldots$ & -55.06 $\pm$ 0.10 \\
183 & 16005980-5334552 & 15.188 & 1.033 & 240.24917 & -53.58200 & $\ldots$ & $\ldots$ & $\ldots$ & $\ldots$ & -38.10 $\pm$ 2.32 \\
189 & 16005475-5335022 & 15.400 & 1.576 & 240.22813 & -53.58394 & $\ldots$ & $\ldots$ & $\ldots$ & $\ldots$ & 78.38 $\pm$ 0.14 \\
191 & 16003979-5329242 & 15.183 & 1.191 & 240.16579 & -53.49006 & 5083 $\pm$ 93 & 4.55 $\pm$ 0.04 & 0.74 $\pm$ 0.04 & 0.18 $\pm$ 0.14 & -31.70 $\pm$ 0.10 \\
208 & 16005846-5335490 & 15.342 & 1.314 & 240.24358 & -53.59694 & $\ldots$ & $\ldots$ & $\ldots$ & $\ldots$ & -55.99 $\pm$ 1.56 \\
209 & 16005739-5332261 & 15.225 & 1.208 & 240.23913 & -53.54058 & $\ldots$ & $\ldots$ & $\ldots$ & $\ldots$ & -68.80 $\pm$ 0.14 \\
212 & 16010451-5332442 & 15.240 & 1.204 & 240.26879 & -53.54561 & $\ldots$ & $\ldots$ & $\ldots$ & $\ldots$ & -50.19 $\pm$ 0.10 \\
213 & 16005402-5331231 & 15.721 & 1.944 & 240.22508 & -53.52308 & 5257 $\pm$ 108 & 3.16 $\pm$ 0.05 & 1.41 $\pm$ 0.03 & -0.10 $\pm$ 0.24 & 13.86 $\pm$ 0.12 \\
214 & 16004351-5333456 & 15.403 & 1.418 & 240.18129 & -53.56267 & $\ldots$ & $\ldots$ & $\ldots$ & $\ldots$ & -31.21 $\pm$ 0.10 \\
216 & 16004226-5335450 & 15.356 & 1.243 & 240.17608 & -53.59583 & $\ldots$ & $\ldots$ & $\ldots$ & $\ldots$ & -68.53 $\pm$ 2.95 \\
217 & 16003389-5330469 & 15.449 & 1.248 & 240.14121 & -53.51303 & $\ldots$ & $\ldots$ & $\ldots$ & $\ldots$ & -48.00 $\pm$ 1.70 \\
219 & 16010922-5330561 & 14.936 & 1.035 & 240.28842 & -53.51558 & $\ldots$ & $\ldots$ & $\ldots$ & $\ldots$ & -21.27 $\pm$ 3.42 \\
222 & 16005026-5333580 & 15.377 & 1.306 & 240.20942 & -53.56611 & $\ldots$ & $\ldots$ & $\ldots$ & $\ldots$ & -3.47 $\pm$ 1.08 \\
225 & 16010322-5332285 & 15.431 & 1.437 & 240.26342 & -53.54125 & $\ldots$ & $\ldots$ & $\ldots$ & $\ldots$ & -51.04 $\pm$ 1.60 \\
226 & 16010444-5330031 & 15.351 & 1.302 & 240.26850 & -53.50086 & $\ldots$ & $\ldots$ & $\ldots$ & $\ldots$ & -68.84 $\pm$ 2.72 \\
247 & 16004890-5328089 & 15.524 & 1.430 & 240.20375 & -53.46914 & $\ldots$ & $\ldots$ & $\ldots$ & $\ldots$ & -16.31 $\pm$ 0.16 \\
252 & 16010967-5332235 & 15.573 & 1.476 & 240.29029 & -53.53986 & $\ldots$ & $\ldots$ & $\ldots$ & $\ldots$ & -55.20 $\pm$ 0.10 \\
255 & 16003791-5331140 & 15.552 & 1.056 & 240.15796 & -53.52056 & $\ldots$ & $\ldots$ & $\ldots$ & $\ldots$ & -44.08 $\pm$ 3.49 \\
261 & 16004556-5332159 & 15.611 & 1.393 & 240.18983 & -53.53775 & 4542 $\pm$ 144 & 4.61 $\pm$ 0.43 & $\ldots$ & 0.15 $\pm$ 0.16 & -24.98 $\pm$ 0.24 \\
262 & 16010496-5332496 & 15.541 & 1.331 & 240.27067 & -53.54711 & 6614 $\pm$ 305 & 4.15 $\pm$ 0.23 & $\ldots$ & -0.04 $\pm$ 0.19 & 7.52 $\pm$ 0.24 \\
276 & 16005205-5335459 & 15.598 & 1.257 & 240.21687 & -53.59608 & $\ldots$ & $\ldots$ & $\ldots$ & $\ldots$ & -35.45 $\pm$ 0.25 \\
279 & 16004926-5333097 & 15.545 & 1.183 & 240.20525 & -53.55269 & $\ldots$ & $\ldots$ & $\ldots$ & $\ldots$ & -91.20 $\pm$ 0.15 \\
303 & 16010676-5333432 & 15.714 & 1.328 & 240.27817 & -53.56200 & $\ldots$ & $\ldots$ & $\ldots$ & $\ldots$ & -37.55 $\pm$ 1.11 \\
304 & 16004058-5329583 & 15.734 & 1.385 & 240.16908 & -53.49953 & $\ldots$ & $\ldots$ & $\ldots$ & $\ldots$ & -66.25 $\pm$ 0.11 \\
314 & 16003652-5330052 & 15.699 & 1.223 & 240.15217 & -53.50144 & $\ldots$ & $\ldots$ & $\ldots$ & $\ldots$ & -80.31 $\pm$ 0.10 \\
315 & 16003754-5333550 & 15.724 & 1.274 & 240.15642 & -53.56528 & 7582 $\pm$ 48 & $\ldots$ & $\ldots$ & $\ldots$ & -31.20 $\pm$ 0.41 \\
331 & 16003799-5334228 & 15.874 & 1.409 & 240.15829 & -53.57300 & 6569 $\pm$ 39 & 4.46 $\pm$ 0.08 & $\ldots$ & 0.45 $\pm$ 0.03 & -4.58 $\pm$ 0.33 \\
344 & 16004804-5333020 & 15.849 & 1.386 & 240.20017 & -53.55056 & $\ldots$ & $\ldots$ & $\ldots$ & $\ldots$ & -14.73 $\pm$ 0.24 \\
353 & 16010155-5328151 & 15.795 & 1.191 & 240.25646 & -53.47086 & $\ldots$ & $\ldots$ & $\ldots$ & $\ldots$ & 458.82 $\pm$ 0.10 \\
366 & 16010826-5331372 & 15.808 & 1.243 & 240.28442 & -53.52700 & 4967 $\pm$ 151 & 4.56 $\pm$ 0.36 & $\ldots$ & 0.04 $\pm$ 0.14 & 5.64 $\pm$ 0.24 \\
370 & 16003334-5329569 & 15.889 & 1.124 & 240.13892 & -53.49914 & 5897 $\pm$ 101 & 4.16 $\pm$ 0.06 & 1.17 $\pm$ 0.03 & -0.18 $\pm$ 0.15 & -28.76 $\pm$ 0.10 \\
373 & 16004375-5335538 & 15.970 & 1.278 & 240.18229 & -53.59828 & $\ldots$ & $\ldots$ & $\ldots$ & $\ldots$ & -33.92 $\pm$ 0.13 \\
381 & 16004932-5332384 & 15.945 & 1.149 & 240.20550 & -53.54400 & 5365 $\pm$ 87 & 4.27 $\pm$ 0.02 & 1.50 $\pm$ 0.07 & 0.09 $\pm$ 0.18 & -24.87 $\pm$ 0.11 \\
382 & 16005577-5329141 & 15.867 & 1.199 & 240.23237 & -53.48725 & $\ldots$ & $\ldots$ & $\ldots$ & $\ldots$ & -86.37 $\pm$ 0.12 \\
384 & 16004196-5335545 & 16.039 & 1.273 & 240.17483 & -53.59847 & $\ldots$ & $\ldots$ & $\ldots$ & $\ldots$ & -25.35 $\pm$ 0.36 \\
446 & 16010350-5329301 & 16.023 & 1.081 & 240.26458 & -53.49169 & 8034 $\pm$ 67 & $\ldots$ & $\ldots$ & $\ldots$ & -45.83 $\pm$ 0.61 \\
464 & 16005229-5333251 & 16.178 & 1.310 & 240.21787 & -53.55697 & $\ldots$ & $\ldots$ & $\ldots$ & $\ldots$ & -33.96 $\pm$ 0.15 \\
476 & 16005961-5331283 & 16.310 & 1.428 & 240.24838 & -53.52453 & $\ldots$ & $\ldots$ & $\ldots$ & $\ldots$ & -67.45 $\pm$ 8.35 \\
492 & 16003558-5329192 & 16.222 & 1.183 & 240.14825 & -53.48867 & $\ldots$ & $\ldots$ & $\ldots$ & $\ldots$ & -75.27 $\pm$ 0.11 \\
494 & 16005700-5330159 & 16.400 & 1.484 & 240.23750 & -53.50442 & 6535 $\pm$ 54 & 3.81 $\pm$ 0.11 & $\ldots$ & 0.44 $\pm$ 0.04 & -47.43 $\pm$ 0.27 \\
497 & 16003142-5330137 & 16.385 & 1.294 & 240.13092 & -53.50381 & 6051 $\pm$ 125 & 4.31 $\pm$ 0.39 & $\ldots$ & -0.01 $\pm$ 0.25 & -35.71 $\pm$ 0.24 \\
499 & 16005117-5330529 & 16.336 & 1.335 & 240.21321 & -53.51469 & $\ldots$ & $\ldots$ & $\ldots$ & $\ldots$ & -65.98 $\pm$ 1.25 \\
514 & 16005793-5329383 & 16.287 & 1.184 & 240.24138 & -53.49397 & 5319 $\pm$ 109 & 4.69 $\pm$ 0.05 & 1.79 $\pm$ 0.03 & -0.19 $\pm$ 0.20 & -37.19 $\pm$ 0.12 \\
526 & 16005684-5329384 & 16.338 & 1.302 & 240.23683 & -53.49400 & $\ldots$ & $\ldots$ & $\ldots$ & $\ldots$ & -22.16 $\pm$ 0.18 \\
530 & 16010748-5333272 & 16.347 & 1.335 & 240.28117 & -53.55756 & $\ldots$ & $\ldots$ & $\ldots$ & $\ldots$ & -17.90 $\pm$ 3.91 \\
540 & 16005040-5331031 & 16.433 & 1.319 & 240.21000 & -53.51753 & $\ldots$ & $\ldots$ & $\ldots$ & $\ldots$ & -112.42 $\pm$ 0.75 \\
542 & 16005288-5331550 & 16.475 & 1.458 & 240.22033 & -53.53194 & $\ldots$ & $\ldots$ & $\ldots$ & $\ldots$ & -12.73 $\pm$ 1.35 \\
543 & 16005553-5330246 & 16.316 & 1.315 & 240.23138 & -53.50683 & 7410 $\pm$ 74 & $\ldots$ & $\ldots$ & $\ldots$ & -4.35 $\pm$ 0.56 \\
560 & 16003787-5331536 & 16.561 & 1.476 & 240.15779 & -53.53156 & 7080 $\pm$ 718 & 4.15 $\pm$ 0.21 & $\ldots$ & 0.12 $\pm$ 0.14 & -54.37 $\pm$ 0.63 \\
572 & 16003903-5333026 & 16.487 & 1.333 & 240.16262 & -53.55072 & $\ldots$ & $\ldots$ & $\ldots$ & $\ldots$ & -69.24 $\pm$ 4.01 \\
584 & 16011001-5331319 & 16.534 & 1.449 & 240.29171 & -53.52553 & 6717 $\pm$ 294 & 4.00 $\pm$ 0.11 & $\ldots$ & 0.15 $\pm$ 0.32 & -29.77 $\pm$ 0.41 \\
585 & 16005947-5328024 & 16.519 & 1.217 & 240.24779 & -53.46733 & 4933 $\pm$ 166 & 4.64 $\pm$ 0.20 & $\ldots$ & 0.04 $\pm$ 0.15 & 10.43 $\pm$ 0.24 \\
591 & 16004161-5333182 & 16.486 & 1.241 & 240.17337 & -53.55506 & 5188 $\pm$ 51 & 4.56 $\pm$ 0.34 & $\ldots$ & 0.02 $\pm$ 0.13 & 42.31 $\pm$ 0.25 \\
621 & 16005269-5333085 & 16.576 & 1.304 & 240.21954 & -53.55236 & 7556 $\pm$ 80 & $\ldots$ & $\ldots$ & $\ldots$ & -66.27 $\pm$ 0.60 \\
651 & 16004658-5334434 & 16.222 & 1.128 & 240.19408 & -53.57872 & $\ldots$ & $\ldots$ & $\ldots$ & $\ldots$ & -14.61 $\pm$ 0.22 \\
659 & 16003500-5334335 & 16.710 & 1.477 & 240.14583 & -53.57597 & 6752 $\pm$ 62 & 4.15 $\pm$ 0.12 & $\ldots$ & 0.45 $\pm$ 0.05 & -47.11 $\pm$ 0.64 \\
685 & 16005470-5334459 & 16.770 & 1.361 & 240.22792 & -53.57942 & 4755 $\pm$ 142 & 4.66 $\pm$ 0.22 & $\ldots$ & -0.14 $\pm$ 0.13 & 48.50 $\pm$ 0.24 \\
709 & 16004956-5329564 & 16.808 & 1.517 & 240.20650 & -53.49900 & 6189 $\pm$ 124 & 3.99 $\pm$ 0.19 & $\ldots$ & 0.02 $\pm$ 0.23 & -40.43 $\pm$ 0.30 \\
747 & 16003312-5332533 & 16.962 & 1.581 & 240.13800 & -53.54814 & 6100 $\pm$ 310 & 4.24 $\pm$ 0.23 & $\ldots$ & 0.30 $\pm$ 0.30 & -88.39 $\pm$ 0.25 \\
779 & 16005584-5333412 & 16.893 & 1.300 & 240.23267 & -53.56144 & 7674 $\pm$ 568 & 4.17 $\pm$ 0.15 & $\ldots$ & 0.02 $\pm$ 0.20 & -49.88 $\pm$ 0.49 \\
787 & 16004146-5328081 & 17.063 & 1.641 & 240.17275 & -53.46892 & 6509 $\pm$ 263 & 4.22 $\pm$ 0.40 & $\ldots$ & 0.32 $\pm$ 0.25 & -32.88 $\pm$ 0.46 \\
828 & 16004452-5329038 & 17.111 & 1.596 & 240.18550 & -53.48439 & 6273 $\pm$ 204 & 4.27 $\pm$ 0.35 & $\ldots$ & 0.08 $\pm$ 0.21 & -34.78 $\pm$ 0.26 \\
878 & 16005429-5329248 & 17.198 & 1.596 & 240.22621 & -53.49022 & 5914 $\pm$ 206 & 4.04 $\pm$ 0.25 & $\ldots$ & -0.25 $\pm$ 0.17 & -90.02 $\pm$ 0.27 \\
907 & 16003886-5335108 & 17.263 & 1.669 & 240.16192 & -53.58633 & 8142 $\pm$ 117 & $\ldots$ & $\ldots$ & $\ldots$ & -20.97 $\pm$ 0.42 \\
920 & 16005372-5328100 & 17.313 & 1.716 & 240.22383 & -53.46944 & 6355 $\pm$ 168 & 3.93 $\pm$ 0.03 & $\ldots$ & 0.20 $\pm$ 0.21 & -37.29 $\pm$ 0.26 \\
982 & 16003991-5335351 & 17.319 & 1.523 & 240.16629 & -53.59308 & 4479 $\pm$ 118 & 4.79 $\pm$ 0.25 & $\ldots$ & -0.03 $\pm$ 0.13 & 48.80 $\pm$ 0.25 \\
990 & 16005659-5327535 & 17.349 & 1.573 & 240.23579 & -53.46486 & 5794 $\pm$ 73 & 4.49 $\pm$ 0.19 & $\ldots$ & -0.21 $\pm$ 0.16 & 25.82 $\pm$ 0.29 \\
992 & 16004925-5334087 & 17.205 & 1.373 & 240.20521 & -53.56908 & 6824 $\pm$ 318 & 4.25 $\pm$ 0.05 & $\ldots$ & 0.29 $\pm$ 0.27 & -26.79 $\pm$ 0.38 \\
996 & 16004583-5330286 & 17.369 & 1.603 & 240.19096 & -53.50794 & 5649 $\pm$ 43 & 4.29 $\pm$ 0.05 & $\ldots$ & 0.00 $\pm$ 0.15 & -0.65 $\pm$ 0.27 \\
1060 & 16010238-5331368 & 17.449 & 1.629 & 240.25992 & -53.52689 & 6369 $\pm$ 193 & 4.56 $\pm$ 0.46 & $\ldots$ & 0.29 $\pm$ 0.23 & -22.50 $\pm$ 0.42 \\
1088 & 16004767-5332334 & 17.543 & 1.760 & 240.19862 & -53.54261 & 5497 $\pm$ 114 & 4.32 $\pm$ 0.10 & $\ldots$ & -0.08 $\pm$ 0.21 & 12.24 $\pm$ 0.27 \\
1111 & 16005412-5328296 & 17.541 & 1.712 & 240.22550 & -53.47489 & 6285 $\pm$ 177 & 4.27 $\pm$ 0.26 & $\ldots$ & 0.10 $\pm$ 0.14 & 13.44 $\pm$ 0.31 \\
1112 & 16005823-5333282 & 17.465 & 1.554 & 240.24263 & -53.55783 & 6253 $\pm$ 389 & 4.17 $\pm$ 0.39 & $\ldots$ & 0.14 $\pm$ 0.33 & -80.74 $\pm$ 0.28 \\
1133 & 16010614-5329010 & 17.518 & 1.607 & 240.27558 & -53.48361 & 5941 $\pm$ 136 & 4.26 $\pm$ 0.05 & $\ldots$ & 0.19 $\pm$ 0.12 & -11.01 $\pm$ 0.25 \\
1139 & 16005380-5336063 & 17.546 & 1.635 & 240.22417 & -53.60175 & 5966 $\pm$ 155 & 4.26 $\pm$ 0.26 & $\ldots$ & 0.02 $\pm$ 0.15 & -40.47 $\pm$ 0.27 \\
1183 & 16005857-5329161 & 17.582 & 1.580 & 240.24404 & -53.48781 & 5882 $\pm$ 115 & 4.10 $\pm$ 0.15 & $\ldots$ & 0.16 $\pm$ 0.15 & -32.19 $\pm$ 0.28 \\
1244 & 16004689-5328513 & 17.640 & 1.516 & 240.19538 & -53.48092 & 6082 $\pm$ 252 & 4.37 $\pm$ 0.22 & $\ldots$ & 0.11 $\pm$ 0.21 & 16.70 $\pm$ 0.27 \\
1416 & 16003805-5331267 & 17.828 & 1.527 & 240.15854 & -53.52408 & 6325 $\pm$ 602 & 4.01 $\pm$ 0.19 & $\ldots$ & 0.02 $\pm$ 0.24 & -68.03 $\pm$ 1.22 \\
1437 & 16005817-5334494 & 17.869 & 1.556 & 240.24238 & -53.58039 & 5274 $\pm$ 109 & 4.71 $\pm$ 0.37 & $\ldots$ & 0.04 $\pm$ 0.12 & 7.75 $\pm$ 0.27 \\
1466 & 16004240-5332325 & 17.976 & 1.768 & 240.17667 & -53.54236 & 5890 $\pm$ 169 & 4.12 $\pm$ 0.18 & $\ldots$ & 0.37 $\pm$ 0.16 & -24.58 $\pm$ 0.26 \\
1475 & 16005947-5330396 & 17.923 & 1.635 & 240.24779 & -53.51100 & 6169 $\pm$ 120 & 4.63 $\pm$ 0.77 & $\ldots$ & -0.01 $\pm$ 0.31 & -54.25 $\pm$ 0.47 \\
1487 & 16010669-5328395 & 17.988 & 1.740 & 240.27787 & -53.47764 & 5238 $\pm$ 115 & 3.99 $\pm$ 0.32 & $\ldots$ & 0.19 $\pm$ 0.29 & -123.55 $\pm$ 0.26 \\
\hline
\end{tabular}
\tablefoot{a. ID numbers and photometry from \citet{Carraro06}.}
\end{table*}

\begin{table*}
\caption{Parameters for Trumpler 23 Targets with UVES Data}
\label{uves_info}
\setlength{\tabcolsep}{0.05in}
\centering
\scriptsize
\begin{tabular}{l c c c c c r c c c c c c c}  
\hline\hline
ID$^{a}$ & GES ID & V$^{a}$ & V-I$^{a}$ & RA & Dec & S/N & RV & \teff & \logg & $\xi$ & [Fe/H] & RV Mem.? \\  & & (mag) & (mag) & (deg.) & (deg.) & & (km s$^{-1}$) & (K) & (cm s$^{-2}$) & (km s$^{-1}$) & (dex) & \\
\hline
52 & 16005168-5332013  & 14.644 & 2.006 & 240.21533 & -53.53369 & 126 & -62.83 $\pm$ 0.57 & 4881 $\pm$ 117 & 2.60 $\pm$ 0.23 & 1.71 $\pm$ 0.16 & 0.17 $\pm$ 0.10 & Y \\
55 & 16003935-5332367  & 14.662 & 2.039 & 240.16396 & -53.54353 & 133 & -60.94 $\pm$ 0.57 & 4796 $\pm$ 123 & 2.57 $\pm$ 0.23 & 1.47 $\pm$ 0.13 & 0.14 $\pm$ 0.10 & Y \\
56 & 16004035-5333047 & 14.731 & 2.089 & 240.16813 & -53.55131 & 125 & -68.64 $\pm$ 0.57 & 4897 $\pm$ 121 & 3.40 $\pm$ 0.26 & 1.26 $\pm$ 0.16 & 0.00 $\pm$ 0.13 & N \\
58 & 16010433-5332336 & 14.743 & 2.017 & 240.26804 & -53.54267 & 66 & -60.53 $\pm$ 0.57 & 4776 $\pm$ 122 & 2.48 $\pm$ 0.24 & 1.68 $\pm$ 0.18 & 0.17 $\pm$ 0.10 & Y \\
59 & 16010770-5329374 & 14.949 & 2.175 & 240.28208 & -53.49372 & 124 & -61.52 $\pm$ 0.57 & 4832 $\pm$ 115 & 2.64 $\pm$ 0.23 & 1.47 $\pm$ 0.12 & 0.16 $\pm$ 0.10 & Y \\
64 & 16005220-5333362 & 14.721 & 1.953 & 240.21750 & -53.56006 & 114 & -62.45 $\pm$ 0.57 & 4917 $\pm$ 116 & 2.63 $\pm$ 0.22 & 1.64 $\pm$ 0.13 & 0.18 $\pm$ 0.10 & Y \\
74 & 16004569-5329177 & 14.769 & 1.867 & 240.19038 & -53.48825 & 120 & -55.40 $\pm$ 0.57 & 5008 $\pm$ 123 & 2.98 $\pm$ 0.23 & 1.45 $\pm$ 0.04 & 0.18 $\pm$ 0.10 & N \\
86 & 16004025-5329439 & 14.992 & 1.958 & 240.16771 & -53.49553 & 111 & -56.47 $\pm$ 0.57 & 4912 $\pm$ 125 & 2.77 $\pm$ 0.24 & 1.51 $\pm$ 0.14 & 0.11 $\pm$ 0.10 & N \\
89 & 16004572-5332095 & 14.948 & 2.031 & 240.19050 & -53.53597 & 84 & -43.27 $\pm$ 0.57 & 4900 $\pm$ 125 & 2.68 $\pm$ 0.26 & 1.53 $\pm$ 0.14 & 0.15 $\pm$ 0.10 & N \\
92 & 16005798-5331476 & 15.086 & 2.077 & 240.24158 & -53.52989 & 83 & -59.95 $\pm$ 0.57 & 4863 $\pm$ 118 & 2.69 $\pm$ 0.24 & 1.46 $\pm$ 0.07 & 0.13 $\pm$ 0.10 & Y \\
97 & 16010639-5331056 & 15.082 & 2.046 & 240.27663 & -53.51822 & 56 & -61.94 $\pm$ 0.57 & 4848 $\pm$ 121 & 2.74 $\pm$ 0.24 & 1.76 $\pm$ 0.17 & 0.14 $\pm$ 0.11 & Y \\
105 & 16010025-5333101 & 15.138 & 2.031 & 240.25104 & -53.55281 & 84 & -60.24 $\pm$ 0.57 & 4884 $\pm$ 119 & 2.79 $\pm$ 0.22 & 1.53 $\pm$ 0.25 & 0.10 $\pm$ 0.10 & Y \\
128 & 16003885-5334507 & 15.320 & 2.148 & 240.16188 & -53.58075 & 80 & -61.15 $\pm$ 0.57 & 4509 $\pm$ 135 & 2.42 $\pm$ 0.24 & 1.49 $\pm$ 0.29 & 0.08 $\pm$ 0.10 & Y \\
131 & 16005072-5335536 & 15.213 & 1.918 & 240.21133 & -53.59822 & 50 & -8.95 $\pm$ 0.57 & 4878 $\pm$ 126 & 2.95 $\pm$ 0.24 & 1.48 $\pm$ 0.05 & 0.19 $\pm$ 0.10 & N \\
137 & 16004312-5330509 & 15.331 & 2.046 & 240.17967 & -53.51414 & 89 & -61.79 $\pm$ 0.57 & 4913 $\pm$ 121 & 2.86 $\pm$ 0.23 & 1.60 $\pm$ 0.25 & 0.11 $\pm$ 0.10 & Y \\
147 & 16004973-5331459 & 15.522 & 2.168 & 240.20721 & -53.52942 & 72 & -10.31 $\pm$ 0.57 & 4449 $\pm$ 126 & 2.43 $\pm$ 0.24 & 1.64 $\pm$ 0.17 & 0.21 $\pm$ 0.09 & N \\
\hline
\end{tabular}
\tablefoot{a. ID numbers and photometry from \citet{Carraro06}.}
\end{table*}

\begin{table*}
\caption{Isochrone Fit Parameters}
\label{iso_params}
\setlength{\tabcolsep}{0.05in}
\centering
\scriptsize
\begin{tabular}{l c c c c c c c}  
\hline\hline
Model & Age & E(V-I) & E(B-V) & (m-M)$_\mathrm{V}$ & d$_{\odot}$ & R$_{\mathrm{GC}}$ & M$_\mathrm{TO}$ \\
 & (Gyr) & (mag) & (mag) & (mag) & (kpc) & (kpc) & (M$_{\odot}$) \\
\hline
PARSEC & 0.80 $\pm$ 0.10 & 1.02$^{+0.14}_{-0.09}$ & 0.82$^{+0.11}_{-0.08}$ & 14.15 $\pm$ 0.20 & 2.10 $\pm$ 0.20 & 6.30 $\pm$ 0.15 & 2.05 \\
BASTI & 0.60 $\pm$ 0.10 & 1.09$^{+0.14}_{-0.09}$ & 0.87$^{+0.11}_{-0.08}$ & 14.50 $\pm$ 0.20 & 2.29 $\pm$ 0.20 & 6.15 $\pm$ 0.15 & 2.17 \\
Dartmouth & 0.65 $\pm$ 0.10 & 1.10$^{+0.14}_{-0.09}$ & 0.88$^{+0.11}_{-0.08}$ & 14.45 $\pm$ 0.20 & 2.21 $\pm$ 0.20 & 6.22 $\pm$ 0.15 & 2.16 \\
Padova \citep{Carraro06} & 0.60 $\pm$ 0.10 & 1.05 $\pm$ 0.05 & 0.84 $\pm$ 0.04 & 14.35 $\pm$ 0.20 & 2.23 $\pm$ 0.20 & 6.20 $\pm$ 0.15 & $\ldots$ \\
Padova \citep{BB07} & 0.9 $\pm$ 0.1 & $\ldots$ & 0.58 $\pm$ 0.03 & 13.19 $\pm$ 0.11 & 1.9 $\pm$ 0.1 & 6.45 $\pm$ 0.08 & $\ldots$ \\
\hline
\end{tabular}
\end{table*}

\begin{table*}
\caption{UVES Target Light and $\alpha$-Element Abundances}
\label{uves_abun1}
\setlength{\tabcolsep}{0.05in}
\centering
\scriptsize
\begin{tabular}{l c c c c c c c c c}
\hline\hline
ID & \ion{C}{I} & \ion{O}{I} & \ion{Na}{I} & \ion{Mg}{I} & \ion{Al}{I} & \ion{Si}{I} & \ion{S}{I} & \ion{Ca}{I} & \ion{Ti}{I} \\
\hline
52   & 8.33 $\pm$ 0.13 & 8.95 $\pm$ 0.25 & 6.92 $\pm$ 0.08 & 7.87 $\pm$ 0.12 & 6.70 $\pm$ 0.06 & 7.72 $\pm$ 0.07 & 7.44 $\pm$ 0.07 & 6.36 $\pm$ 0.08 & 4.98 $\pm$ 0.08 \\
55   & 8.44 $\pm$ 0.13 & 8.91 $\pm$ 0.14 & 6.82 $\pm$ 0.06 & 7.87 $\pm$ 0.12 & 6.67 $\pm$ 0.07 & 7.62 $\pm$ 0.07 & 7.39 $\pm$ 0.07 & 6.43 $\pm$ 0.09 & 4.98 $\pm$ 0.08 \\
58   & 8.45 $\pm$ 0.13 & 8.80 $\pm$ 0.11 & 6.78 $\pm$ 0.05 & 7.91 $\pm$ 0.12 & 6.70 $\pm$ 0.07 & 7.60 $\pm$ 0.07 & 7.35 $\pm$ 0.07 & 6.36 $\pm$ 0.08 & 4.95 $\pm$ 0.08 \\
59   & 8.28 $\pm$ 0.13 & 8.79 $\pm$ 0.33 & 6.90 $\pm$ 0.08 & 7.86 $\pm$ 0.12 & 6.65 $\pm$ 0.07 & 7.71 $\pm$ 0.07 & 7.41 $\pm$ 0.07 & 6.48 $\pm$ 0.08 & 5.06 $\pm$ 0.08 \\
64   & 8.40 $\pm$ 0.13 & 8.92 $\pm$ 0.23 & 6.80 $\pm$ 0.06 & 7.88 $\pm$ 0.12 & 6.68 $\pm$ 0.07 & 7.64 $\pm$ 0.07 & 7.38 $\pm$ 0.07 & 6.43 $\pm$ 0.08 & 4.93 $\pm$ 0.09 \\
92   & 8.34 $\pm$ 0.13 & 8.81 $\pm$ 0.12 & 6.85 $\pm$ 0.06 & 7.82 $\pm$ 0.12 & 6.65 $\pm$ 0.07 & 7.66 $\pm$ 0.07 & 7.42 $\pm$ 0.07 & 6.36 $\pm$ 0.08 & 4.92 $\pm$ 0.09 \\
97   & 8.40 $\pm$ 0.13 & 8.85 $\pm$ 0.12 & 6.79 $\pm$ 0.05 & 7.78 $\pm$ 0.12 & 6.71 $\pm$ 0.07 & 7.66 $\pm$ 0.07 & 7.46 $\pm$ 0.07 & 6.33 $\pm$ 0.08 & 4.96 $\pm$ 0.08 \\
105 & 8.25 $\pm$ 0.13 & 8.80 $\pm$ 0.12 & 6.81 $\pm$ 0.05 & 7.67 $\pm$ 0.12 & 6.70 $\pm$ 0.06 & 7.67 $\pm$ 0.07 & 7.43 $\pm$ 0.07 & 6.42 $\pm$ 0.08 & 4.94 $\pm$ 0.08 \\
128 & 8.47 $\pm$ 0.13 & 8.84 $\pm$ 0.12 & 6.61 $\pm$ 0.08 & 7.95 $\pm$ 0.12 & 6.71 $\pm$ 0.07 & 7.63 $\pm$ 0.07 & 7.45 $\pm$ 0.07 & 6.36 $\pm$ 0.09 & 5.02 $\pm$ 0.08 \\
137 & 8.24 $\pm$ 0.13 & 8.93 $\pm$ 0.12 & 6.78 $\pm$ 0.06 & 7.92 $\pm$ 0.12 & 6.69 $\pm$ 0.06 & 7.69 $\pm$ 0.07 & 7.42 $\pm$ 0.07 & 6.42 $\pm$ 0.08 & 5.05 $\pm$ 0.09 \\
\hline
\end{tabular}
\end{table*}

\begin{table*}
\caption{UVES Target Fe-Peak Abundances}
\label{uves_abun2}
\setlength{\tabcolsep}{0.05in}
\centering
\scriptsize
\begin{tabular}{l c c c c c c c c}
\hline\hline
ID & \ion{Sc}{I} &  \ion{V}{I} & \ion{Cr}{I} & \ion{Mn}{I} & \ion{Fe}{I} & \ion{Fe}{II} & \ion{Co}{I} & \ion{Ni}{I} \\
\hline
52 & 3.24 $\pm$ 0.10 & 4.09 $\pm$ 0.09 & 5.72 $\pm$ 0.12 & 5.46 $\pm$ 0.12 & 7.57 $\pm$ 0.10 & 7.66 $\pm$ 0.09 & 5.05 $\pm$ 0.11 & 6.38 $\pm$ 0.09 \\
55 & 3.42 $\pm$ 0.08 & 4.10 $\pm$ 0.09 & 5.73 $\pm$ 0.13 & 5.39 $\pm$ 0.12 & 7.61 $\pm$ 0.10 & 7.65 $\pm$ 0.09 & 5.06 $\pm$ 0.10 & 6.40 $\pm$ 0.10 \\
58 & 3.35 $\pm$ 0.08 & 4.07 $\pm$ 0.08 & 5.66 $\pm$ 0.10 & 5.39 $\pm$ 0.11 & 7.56 $\pm$ 0.10 & 7.55 $\pm$ 0.09 & 5.02 $\pm$ 0.10 & 6.33 $\pm$ 0.10 \\
59 & 3.36 $\pm$ 0.09 & 4.15 $\pm$ 0.09 & 5.78 $\pm$ 0.11 & 5.36 $\pm$ 0.12 & 7.63 $\pm$ 0.10 & 7.70 $\pm$ 0.09 & 5.10 $\pm$ 0.10 & 6.44 $\pm$ 0.10 \\
64 & 3.24 $\pm$ 0.10 & 4.05 $\pm$ 0.09 & 5.66 $\pm$ 0.12 & 5.53 $\pm$ 0.14 & 7.57 $\pm$ 0.09 & 7.63 $\pm$ 0.09 & 5.03 $\pm$ 0.10 & 6.37 $\pm$ 0.10 \\
92 & 3.29 $\pm$ 0.09 & 4.05 $\pm$ 0.09 & 5.67 $\pm$ 0.11 & 5.36 $\pm$ 0.14 & 7.55 $\pm$ 0.10 & 7.58 $\pm$ 0.09 & 5.00 $\pm$ 0.10 & 6.30 $\pm$ 0.11 \\
97 & 3.04 $\pm$ 0.07 & 4.04 $\pm$ 0.08 & 5.66 $\pm$ 0.11 & 5.38 $\pm$ 0.11 & 7.50 $\pm$ 0.10 & 7.58 $\pm$ 0.09 & 5.06 $\pm$ 0.09 & 6.31 $\pm$ 0.10 \\
105 & 3.27 $\pm$ 0.09 & 4.09 $\pm$ 0.09 & 5.73 $\pm$ 0.12 & 5.39 $\pm$ 0.13 & 7.55 $\pm$ 0.09 & 7.63 $\pm$ 0.08 & 4.98 $\pm$ 0.10 & 6.33 $\pm$ 0.10 \\
128 & 3.28 $\pm$ 0.09 & 4.13 $\pm$ 0.09 & 5.72 $\pm$ 0.11 & 5.41 $\pm$ 0.13 & 7.58 $\pm$ 0.10 & 7.52 $\pm$ 0.10 & 5.16 $\pm$ 0.10 & 6.47 $\pm$ 0.10 \\
137 & 3.37 $\pm$ 0.10 & 4.13 $\pm$ 0.09 & 5.76 $\pm$ 0.11 & 5.46 $\pm$ 0.13 & 7.58 $\pm$ 0.10 & 7.64 $\pm$ 0.08 & 5.04 $\pm$ 0.10 & 6.35 $\pm$ 0.12 \\
\hline
\end{tabular}
\end{table*}

\begin{table*}
\caption{UVES Target Neutron-Capture Element Abundances}
\label{uves_abun3}
\setlength{\tabcolsep}{0.05in}
\centering
\scriptsize
\begin{tabular}{l c c c c c c c c}
\hline\hline
ID & \ion{Y}{II} &  \ion{Zr}{I} & \ion{Mo}{I} & \ion{Ba}{II} & \ion{La}{II} & \ion{Ce}{II} & \ion{Nd}{II} & \ion{Eu}{II} \\
\hline
52 & 2.10 $\pm$ 0.11 & 2.58 $\pm$ 0.15 & 1.71 $\pm$ 0.10 & 2.32 $\pm$ 0.15 & 0.96 $\pm$ 0.15 & 1.77 $\pm$ 0.15 & 1.34 $\pm$ 0.17 & 0.68 $\pm$ 0.17 \\
55 & 2.14 $\pm$ 0.11 & 2.56 $\pm$ 0.15 & 1.77 $\pm$ 0.10 & 2.28 $\pm$ 0.15 & 1.06 $\pm$ 0.15 & 1.62 $\pm$ 0.14 & 1.45 $\pm$ 0.17 & 0.76 $\pm$ 0.10 \\
58 & 2.09 $\pm$ 0.11 & 2.54 $\pm$ 0.14 & 1.71 $\pm$ 0.10 & 2.11 $\pm$ 0.14 & 0.87 $\pm$ 0.15 & 1.72 $\pm$ 0.15 & 1.39 $\pm$ 0.18 & 0.72 $\pm$ 0.10 \\
59 & 2.06 $\pm$ 0.10 & 2.76 $\pm$ 0.13 & 1.72 $\pm$ 0.10 & 2.48 $\pm$ 0.15 & 0.83 $\pm$ 0.15 & 1.51 $\pm$ 0.13 & 1.60 $\pm$ 0.15 & 0.72 $\pm$ 0.15 \\
64 & 2.16 $\pm$ 0.11 & 2.57 $\pm$ 0.15 & 1.77 $\pm$ 0.10 & 2.86 $\pm$ 0.15 & 0.93 $\pm$ 0.15 & 1.57 $\pm$ 0.14 & 1.36 $\pm$ 0.16 & 0.65 $\pm$ 0.17 \\
92 & 2.13 $\pm$ 0.12 & 2.58 $\pm$ 0.13 & 1.75 $\pm$ 0.10 & 2.32 $\pm$ 0.15 & 0.90 $\pm$ 0.15 & 1.47 $\pm$ 0.13 & 1.42 $\pm$ 0.16 & 0.59 $\pm$ 0.16 \\
97 & 2.11 $\pm$ 0.10 & 2.57 $\pm$ 0.14 & 1.98 $\pm$ 0.10 & 1.98 $\pm$ 0.14 & 0.85 $\pm$ 0.15 & 1.63 $\pm$ 0.13 & 1.58 $\pm$ 0.15 & 0.79 $\pm$ 0.10 \\
105 & 2.18 $\pm$ 0.12 & 2.71 $\pm$ 0.14 & 1.75 $\pm$ 0.11 & 2.31 $\pm$ 0.14 & 0.86 $\pm$ 0.15 & 1.59 $\pm$ 0.14 & 1.43 $\pm$ 0.19 & 0.69 $\pm$ 0.16 \\
128 & 2.14 $\pm$ 0.11 & 2.54 $\pm$ 0.14 & 1.78 $\pm$ 0.10 & 2.23 $\pm$ 0.14 & 0.90 $\pm$ 0.15 & 1.44 $\pm$ 0.21 & 1.76 $\pm$ 0.15 & 0.66 $\pm$ 0.10 \\
137 & 2.20 $\pm$ 0.12 & 2.73 $\pm$ 0.13 & 1.78 $\pm$ 0.10 & 2.42 $\pm$ 0.14 & 1.01 $\pm$ 0.15 & 1.64 $\pm$ 0.15 & 1.65 $\pm$ 0.18 & 0.62 $\pm$ 0.17 \\
\hline
\end{tabular}
\end{table*}

\begin{table*}
\caption{Cluster Average Element Abundances}
\label{avg_abun}
\setlength{\tabcolsep}{0.05in}
\centering
\scriptsize
\begin{tabular}{l c c r r r}
\hline\hline
 & G07 & [X/Fe]$^{\mathrm{a}}$ & $\sigma$X & $\sigma$X & $\delta$X \\
Species & solar & (dex) & (stdev) & (err. mean) & (med. st. error) \\
\hline
\ion{C}{I}    & 8.39 & -0.15 & 0.09 & 0.03 & 0.13 \\
\ion{O}{I}    & 8.66 & 0.08 & 0.07 & 0.02 & 0.12 \\
\ion{Na}{I}$^{\mathrm{b}}$ & 6.17 & 0.42 & 0.08 & 0.03 & 0.06 \\
\ion{Mg}{I} & 7.53 & 0.20 & 0.07 & 0.02 & 0.12 \\
\ion{Al}{I}   & 6.37 & 0.20 & 0.05 & 0.02 & 0.07 \\
\ion{Si}{I}   & 7.51 & 0.03 & 0.05 & 0.02 & 0.07 \\
\ion{S}{I}    & 7.14 & 0.16 & 0.06 & 0.02 & 0.07 \\
\ion{Ca}{I} & 6.31 & -0.04 & 0.03 & 0.01 & 0.08 \\
\ion{Sc}{I}  & 3.17 & 0.00 & 0.08 & 0.02 & 0.09 \\
\ion{Ti}{I}   & 4.90 & -0.04 & 0.04 & 0.01 & 0.08 \\
\ion{V}{I}    & 4.00 & -0.03 & 0.02 & 0.01 & 0.09 \\
\ion{Cr}{I}   & 5.64 & -0.05 & 0.03 & 0.01 & 0.11 \\
\ion{Mn}{I}  & 5.39 & -0.10 & 0.06 & 0.02 & 0.13 \\
\ion{Fe}{I}   & 7.45 & 7.57 & 0.04 & 0.01 & 0.10 \\
\ion{Fe}{II}  & 7.45 & 7.61 & 0.06 & 0.02 & 0.09 \\
\ion{Co}{I}  & 4.92 & 0.01 & 0.05 & 0.02 & 0.10 \\
\ion{Ni}{I}   & 6.23 & 0.02 & 0.04 & 0.01 & 0.10 \\
\ion{Y}{II}   & 2.21 & -0.24 & 0.07 & 0.02 & 0.11 \\
\ion{Zr}{I}   & 2.58 & -0.09 & 0.08 & 0.02 & 0.14 \\
\ion{Mo}{I} & 1.92 & -0.27 & 0.10 & 0.03 & 0.10 \\
\ion{Ba}{II} & 2.17 & -0.03 & 0.08 & 0.02 & 0.15 \\
\ion{La}{II} & 1.13 & -0.38 & 0.08 & 0.03 & 0.15 \\
\ion{Ce}{II} & 1.70 & -0.27 & 0.11 & 0.03 & 0.14 \\
\ion{Nd}{II} & 1.45 & -0.12 & 0.16 & 0.05 & 0.17 \\
\ion{Eu}{II} & 0.52 & 0.00 & 0.08 & 0.02 & 0.16 \\
\hline
\end{tabular}
\tablefoot{a. Ratios for neutral species are calculated relative to Fe I, and singly ionized species relative to Fe II. Cluster averages use \citet{Grevesse07} solar abundances.}
\tablefoot{b. Tr 23 [Na/Fe] includes NLTE corrections.}
\end{table*}

\end{document}